\documentclass[AMA,LATO1COL]{WileyNJD-v2}
\usepackage{caption}
\usepackage{subcaption}
\usepackage{siunitx}
\usepackage{multirow}
\usepackage{placeins}
\usepackage{lineno}

\articletype{Article Type}

\received{}
\revised{}
\accepted{}

\begin{document}

\title{Modeling the effect of wind speed and direction shear on utility-scale wind turbine power production}

\author[1]{Storm A. Mata}

\author[2]{Juan Jos{\'e} Pena Mart{\'i}nez}
\author[2]{Jes{\'u}s Bas Quesada}
\author[2]{Felipe Palou Larra\~{n}aga}
\author[3]{Neeraj Yadav}
\author[3]{Jasvipul S. Chawla}
\author[4]{Varun Sivaram}

\author[1]{Michael F. Howland}

\authormark{S. A. Mata {\it et al.}}

\address[1]{\orgdiv{Department of Civil and Environmental Engineering}, \orgname{Massachusetts Institute of Technology}, \orgaddress{\state{Massachusetts}, \country{USA}}}

\address[2]{Siemens Gamesa Renewable Energy Innovation \& Technology, Sarriguren, Navarra, Spain, 31621}
\address[3]{ReNew Power Private Limited, Gurugram-122009, Haryana, India}
\address[4]{{\O}rsted, 1225 New York Avenue NW, Suite 550B Washington, DC 20005}

\corres{M. F. Howland, Department of Civil and Environmental Engineering, Massachusetts Institute of Technology, Cambridge, Massachusetts 02139, USA. \email{mhowland@mit.edu}}

\abstract[Summary]{Wind speed and direction variations across the rotor affect power production. As utility-scale turbines extend higher into the atmospheric boundary layer (ABL) with larger rotor diameters and hub heights, they increasingly encounter more complex wind speed and direction variations. We assess three models for power production that account for wind speed and direction shear. Two are based on actuator disc representations and the third is a blade element representation. We also evaluate the predictions from a standard power curve model that has no knowledge of wind shear. The predictions from each model, driven by wind profile measurements from a profiling LiDAR, are compared to concurrent power measurements from an adjacent utility-scale wind turbine. In the field measurements of the utility-scale turbine, discrete combinations of speed and direction shear induce changes in power production of -19\% to +34\% relative to the turbine power curve for a given hub height wind speed. Positive speed shear generally corresponds to over-performance and positive direction shear to under-performance, relative to the power curve. Overall, the blade element model produces both higher correlation and lower error relative to the other models, but its quantitative accuracy depends on induction and controller sub-models. To further assess the influence of complex, non-monotonic wind profiles, we also drive the models with best-fit power law wind speed profiles and linear wind direction profiles. These idealized inputs produce qualitative and quantitative differences in power predictions from each model, demonstrating that time-varying, non-monotonic wind shear affects wind power production.}


\jnlcitation{\cname{%
\author{Mata S, Pena Mart{\'i}nez JJ, Bas Quesada J, Palou Larra\~{n}aga F, Yadav N, Chawla JS, Sivaram V, Howland MF.}, \ctitle{Modeling the effect of wind speed and direction shear on utility-scale wind turbine power production}, \cjournal{Wind Energy}. (\cyear{2023});  \cvol{vol. XXX}.}}

\maketitle

\section{Introduction}\label{sec:intro}

Mitigation of anthropogenic climate change depends on the decarbonization of carbon-intensive industries, in particular, electricity generation. The share of wind energy generation in global electricity production has grown rapidly in the preceding two decades, undergoing a factor of 35x growth in total installed capacity over that time. Concurrently, the physical dimensions of utility-scale turbines have also grown at a rapid rate. In the United States, the current average hub height (94 m) and rotor diameter (127 m) of land-based turbines are 66\% and 164\% larger, respectively, than those 20 years ago.~\cite{Land_Based_2022} For off-shore turbines, the average hub height (108 m) and rotor diameter (158 m) are 15\% and 24\% larger, respectively, than their land-based counterparts.~\cite{Offshore_2022} The growth in the physical size of wind turbines has corresponded to an increase in their nameplate capacity, as well, with the average generating capacity of onshore turbines nearly tripling from around 1 MW to around 3 MW, and for off-shore turbines nearly quadrupling from around 2 MW to around 9 MW.~\cite{Land_Based_2022, Offshore_2022} These advances in wind energy technology portend future challenges in turbine power modeling as larger physical dimensions correspond to an increased potential for experiencing complex wind conditions that affect power production.~\cite{Ryu_2022}

As the size of both land-based and off-shore turbines grows, they reach farther into the atmospheric boundary layer (ABL) where they may experience more complex wind profiles.~\cite{Veers_2019} In particular, a characteristic feature of ABL flows is vertical wind shear. In this study, we define speed shear as the change in the wind speed profile as a function of height, and direction shear as a change in the wind direction profile with height. Given the presence of the surface, which limits vertical length scales, vertical wind shear tends to dominate horizontal wind shear in the ABL except in very complex terrain~\cite{Stull_ch1,lange2017wind} and spatially heterogeneous wake interactions.~\cite{liew2020analytical} Therefore, this study focuses only on vertical wind shear. However, we note that the influence of horizontal wind shear may be modeled similarly to that of vertical wind shear if information about horizontal wind shear is available. Shear is important to consider in modeling turbine power production because changes in wind speed and direction alter the momentum and kinetic energy flux through the rotor.~\cite{Lundquist_2020} Moreover, given its dependence on atmospheric characteristics such as Coriolis forces, friction, stability, and large-scale forcings, the degree of shear varies both as a function of time and geographic location.~\cite{Wyngaard_ch9,Debnath_2023} The inherently site-specific nature of wind shear motivates the need for an accurate parametric model that incorporates its effect on wind power production, beyond assessment of empirical trends.

Several factors are know to influence the degree of shear present in the ABL. At a given location, the degree of wind shear varies in time depending on the stability in the ABL.~\cite{Walter_2009,howland2022optimal} For instance, Van Ulden and Holtslag~\cite{Van_Ulden1985} analyzed a set of measurements of wind speed and temperature profiles in the atmosphere taken across several decades and found that the magnitude of direction shear, in particular, is strongly correlated with stratification. Further, Pe\~{n}a et al.~\cite{Pena2014} found similar results when analyzing a set of LiDAR and meterological mast data in which both speed and direction shear tend to be stronger during times of stable stratification than unstable or neutral stratification. Further, wind shear depends on the particular geographic location of interest. Since wind direction shear primarily manifests from Coriolis effects in the Ekman spiral, its magnitude inherently depends on the latitude.~\cite{Wyngaard_ch9} Lindvall and Svensson\cite{Lindvall2019} found that local topography influences the degree of shear that may develop in a given location, with coastal regions generally seeing lower degrees of direction shear than inland regions. Shu et al.~\cite{Shu_2020} also found that in these coastal regions, winds flowing from land towards open water tend to contain larger degrees of direction shear than winds that flow from open water onto land. Finally, Wharton and Lundquist\cite{Wharton_2012a,Wharton_2012b} observed that even in stably stratified flows where high direction shear would normally be expected, topography that tends to form channeled flow, such as valleys, appeared to mitigate the frequency of shear.

Given the inherently site-specific nature of wind shear and wind turbine design, drawing definitive conclusions with respect to the impact of shear on turbine power production has remained elusive. Wharton and Lundquist\cite{Wharton_2012a,Wharton_2012b} found that at one site, stable stratification and high shear tended to correlate with higher turbine efficiency, whereas at another site, they found the opposite to be true. Additional studies by St. Martin et al.~\cite{StMartin_2016} and Venderwende and Lundquist\cite{Vanderwende_2012} found different effects of shear on power production relative to the operating regime of the turbines at their respective locations, with stable conditions generally corresponding to increased power production near rated wind speeds, and decreased power production below rated wind speeds. Murphy et al.~\cite{Murphy_2020} found similar results, indicating that direction shear occurring at wind speeds below rated had generally statistically insignificant effects on power production. They found furthermore that the degree of speed shear was only weakly correlated with changes in power production. Howland et al.~\cite{Howland_2020,howland2022collective} found that wind speed and direction shear modify the power production of wind turbines in yaw misalignment, and proposed a blade element model to account for the observed influence. Finally, Sanchez Gomez and Lundquist\cite{MSG_2020} analyzed the effect of both speed and direction shear, and found that large direction shear and small speed shear tended to correspond to decreased power production, while large speed shear and small direction shear tended to result in greater power production. The empirical results discussed above are highly varied given the site-specific trends in wind characteristics and topography in each study. The general lack of agreement in the results across studies that consider shear in relation to turbine performance suggests a few conclusions: 1) the topography, stability, and resulting characteristic wind conditions at a given location are strongly related to the observed trends in power production, and 2) the method of analysis, as well as the specific wind profile and wind turbine characteristics considered, affect the observed trends in power production.

There have been several models proposed for turbine power production to account for variations in wind speed and direction over the turbine rotor area, however, there exists no clear consensus as to which is most accurate. In this study, we compare three parametric models that take in arbitrary wind speed and direction profiles as inputs, and we compare their power predictions to field measurements to quantify their accuracy. Further, to identify effects of site-specific wind characteristics, as well as the general effect of complexity in wind profiles on model outputs, we drive the models with two input classes separately: 1) best-fit canonical ABL profiles in which wind speed follows a power law relationship as a function of height and direction shear is linear over the rotor, and 2) wind speed and direction observations measured by a profiling LiDAR. This analysis seeks to demonstrate the potential variation produced in the model outputs when smooth, monotonic wind profiles are assumed in place of finite time-averaged, empirical wind measurements that permit non-monotonic shear. 

The organization of this study is as follows: in Sec.~\ref{sec:turbine_models} we provide a description of the turbine models considered in this study, including a discussion of the inputs required for each, the mechanism by which each accounts for shear, and an overview of their numerical implementation. Sec.~\ref{sec:Exp_Design} provides a description of the experimental design in which the supervisory control and data acquisition (SCADA) power measurements and LiDAR wind profile measurements were recorded. This section also includes a justification of the selection of two distinct input classes used to drive the models. Sec.~\ref{sec:Results} includes a quantitative comparison of each of the models and their accuracy in predicting the observed SCADA power by considering both correlation and overall error. We also discuss the qualitative trends observed in power production in both the SCADA power measurements and the model predictions, for both the LiDAR wind profile inputs and the canonical ABL profile inputs. Finally, Sec.~\ref{sec:Discussion} discusses the implications of the observed results and the subsequent future research directions motivated by these findings.

\section{Wind turbine models}\label{sec:turbine_models}

The following provides a description of each of the four models included in this study, along with a brief discussion on the mechanism by which each model considers variations in wind speed and direction. Table \ref{Table:Model_Summary} summarizes the inputs required for each model.

\subsection{Hub height wind speed model}\label{sec:HH_model}

The power available to a wind turbine is given by the kinetic energy flux through the rotor disc,~\cite{Manwell_Ch2} expressed as 

\begin{equation}\label{eq:P_Aero}
P_{\text{Aero}} = \frac{1}{2} \rho A_d U^3, 
\end{equation}

\noindent where $\rho$ is the density of air, $A_d$ is the area of the rotor disc, and $U$ is the speed of the air. Following from Eq.~\ref{eq:P_Aero}, we introduce the commonly used hub height wind speed model, given by

\begin{equation}\label{eq:P_HH}
P_{\text{HH}} = \frac{1}{2} \rho A_d C_P U(z_h)^3,
\end{equation}

\noindent where $C_P$ is the coefficient of power and $z_h$ is the hub height for a given turbine. Eq.~\ref{eq:P_HH} is one of two methods specified in IEC Standard 61400-12-1, which provides a standardized method for wind resource assessment across different geographic locations with different site-specific wind conditions.~\cite{Lydia_2013,Lydia_2014} A feature of Eq.~\ref{eq:P_HH} that bears emphasizing is that it considers wind speed at only a single point ($z=z_h$). Thus, to be strictly accurate, Eq.~\ref{eq:P_HH} only applies in cases of uniform inflow where wind speed and direction are constant over the rotor area (i.e., no speed or direction shear), or where the effect of wind speed and direction shear is empirically incorporated into the coefficient of power $C_P$, which is challenging to apply in a generalized, parametric manner. Consequently, the hub height wind speed model is functionally incapable of directly modeling the dependence of wind power production on wind shear. Therefore, the results of the hub height wind speed model serve as a benchmark against which the results for the other models considered in this study are compared.

\subsection{Rotor-equivalent wind speed model}\label{REWS_model}

\begin{center}
\begin{table*}[t]%
\caption{Overview of model inputs.\label{Table:Model_Summary}}
\centering
\begin{tabular*}{500pt}{@{\extracolsep\fill}lccccc@{\extracolsep\fill}}
\toprule
&\multicolumn{2}{@{}c@{}}{\textbf{Wind}} & \multicolumn{3}{@{}c@{}}{\textbf{Turbine}} \\\cmidrule{2-3}\cmidrule{4-6}
\textbf{Model} & \textbf{Speed} & \textbf{Direction} & \textbf{Integrated} & \textbf{Airfoils} & \textbf{Rotor} \\
\midrule
Hub height wind speed       & $U(z_h)$    & \textendash       & $C_P$       & \textendash           & $A_d$                   \\
Rotor-equivalent wind speed & $U(r,\psi)$ & $\Lambda(r,\psi)$ & $C_P$       & \textendash           & $A_d, \zeta$            \\
Rotor-equivalent power      & $U(r,\psi)$ & $\Lambda(r,\psi)$ & $C_P$       & \textendash           & $A_d, \zeta$            \\
Blade element               & $U(r,\psi)$ & $\Lambda(r,\psi)$ & \textendash & $C_L, C_D, c, \theta$ & $A_d, B, \zeta, \Omega$ \\
\bottomrule
\end{tabular*}
\begin{tablenotes}
\item A summary of the inputs to each model. Each model takes in the rotor disc area $A_d$ and the wind speed $U$ at either hub height $z_h$ or as a function of the local radial position $r$ and azimuth angle $\psi$ over the rotor area. The rotor-equivalent wind speed, rotor-equivalent power, and blade element models take in the wind direction $\Lambda$ over the rotor area and the nacelle heading $\zeta$. In Sec.~\ref{sec:turbine_models}, the wind direction and nacelle heading characterize the local wind misalignment angle $\gamma_z(r,\psi) = \Lambda(r,\psi) - \zeta$ over the rotor area. The blade element model also takes in the rotor angular velocity $\Omega$, the number of blades $B$, the blade pitch angle $\theta$, the lift and drag coefficients, $C_L$ and $C_D$, and the chord length $c$ for each blade node. The hub height, rotor-equivalent wind speed, and rotor-equivalent power models take the turbine coefficient of power $C_P$, integrated over the rotor area.
\end{tablenotes}
\end{table*}
\end{center}

This section summarizes the method introduced by Wagner et al.~\cite{Wagner_2009} to account for wind speed variations over the turbine rotor disc area and further expanded by Choukulkar et al.~\cite{Choukulkar_2016} to account for wind direction variations. The form of the rotor-equivalent wind speed (REWS) model is similar to the hub height wind speed model in that turbine power is calculated from the kinetic energy flux through the rotor disc area. However, the REWS model samples the wind field at multiple locations over the rotor disc, thereby accounting for shear. Wagner et al.~\cite{Wagner_2009} defined the rotor-equivalent wind speed as

\begin{equation}\label{eq:U_REWS_Wagner}
U_{eq} = \frac{1}{A_d}\sum_i U_i \cdot A_i,
\end{equation}

\noindent where the rotor disc area is divided into a series of horizontal segments. $U_i$ is the wind speed measured in the $i$th segment and $A_i$ is the corresponding area of that segment. In this way, $U_{eq}$ is a weighted average of the wind speed measurements taken in each segment of the rotor. Choukulkar et al.~\cite{Choukulkar_2016} subsequently expanded on Eq.~\ref{eq:U_REWS_Wagner} by incorporating information about variations in direction over the rotor by computing the rotor-normal wind component with a cosine projection of the incident wind vectors. This requires computing the local wind misalignment angle over the rotor area, given by

\begin{equation}\label{eq:gamma}
\gamma_z(r,\psi) = \Lambda(r,\psi) - \zeta,
\end{equation}

\noindent where $\Lambda$ is the wind direction and $\zeta$ is the turbine nacelle heading. We express the local misalignment angle $\gamma_z$ and the wind direction $\Lambda$ as functions of the local radial position $r$ measured from the rotor center and the azimuth angle $\psi$. The geometry is shown in Fig.~\ref{fig:BE_Diagram}(a). This change in the coordinate system is chosen for convenience in computing the area average of the normal velocity components. Whereas discretizing the rotor in Cartesian sectors requires computing the area of each individual sector, performing the integration in polar coordinates negates this step. Given this modification, the form of the rotor-equivalent wind speed used in this study is given by  

\begin{equation}\label{eq:U_REWS}
U_{\text{REWS}} = \frac{1}{A_d} \int_0^{2\pi} \int_{0}^{R}U(r,\psi)\text{cos}\left[\gamma_z(r,\psi)\right] r\ \text{d}r\ \text{d}\psi.
\end{equation}

\noindent Replacing the hub height wind speed $U(z_h)$ in Eq.~\ref{eq:P_HH} with the rotor-equivalent wind speed $U_{\text{REWS}}$ yields the rotor-equivalent wind speed model, expressed as

\begin{equation}\label{eq:P_REWS}
P_{\text{REWS}} = \frac{1}{2} \rho A_d C_P\left[\frac{1}{A_d} \int_0^{2\pi} \int_{0}^{R}U(r,\psi)\text{cos}\left[\gamma_z(r,\psi)\right] r\ \text{d}r\ \text{d}\psi \right]^3.
\end{equation}

Through simulations of turbine power production using both Eq.~\ref{eq:P_HH} and Eq.~\ref{eq:P_REWS}, Wagner et al.~\cite{Wagner_2009,Wagner_2011} found that the REWS model generally showed higher correlation with measured turbine power when input wind conditions contained large variations in speed and direction. Since it was introduced, the REWS model has been incorporated into studies on a variety of wind energy-related topics, including theoretical studies on the effect of turbulence on power production (Clack et al.~\cite{Clack_2016}), mesoscale modeling and wind farm parameterization (Redfern et al.~\cite{Redfern2019}), and further modifications to the form of the REWS model and their effect on predicted turbine power production (Murphy et al.~\cite{Murphy_2020}).

\subsection{Rotor-equivalent power model}\label{sec:REP_model}

This section summarizes an alternative approach, also described by Wagner et al.~\cite{Wagner_2009} and Choukulkar et al.~\cite{Choukulkar_2016}, for estimating power production from arbitrary wind speed and direction inputs. Whereas Eq.~\ref{eq:P_REWS} estimates turbine power production from the kinetic energy flux produced by a rotor-average wind speed, it is likewise possible to compute the average kinetic energy flux incident on the rotor directly. To accomplish this requires a single modification to Eq.~\ref{eq:P_REWS}. While the REWS model computes $U_{\text{REWS}}$, which is then cubed when substituted into Eq.~\ref{eq:P_HH}, the rotor-equivalent power (REP) model moves the cube inside the area integrals, producing a single value of rotor-equivalent wind speed-cubed. In the REWS model, $U_{\text{REWS}}$ replaces the $U(z_h)$ term in Eq.~\ref{eq:P_HH}, and in the REP model, the rotor-equivalent wind speed-cubed replaces the $U(z_h)^3$ term, yielding 

\begin{equation}\label{eq:P_REP}
P_{\text{REP}} = \frac{1}{2} \rho C_P\int_0^{2\pi} \int_{0}^{R} \left[ U(r,\psi)\text{cos}\left[\gamma_z(r,\psi)\right]  \right]^3 r\ \text{d}r \ \text{d}\psi.
\end{equation}

\noindent From the formula above, it is apparent that if the constant terms outside the integral were moved inside, Eq.~\ref{eq:P_REP} would produce a rotor-averaged value with units of power, hence the rotor-equivalent power label applied in this study. 

\subsection{Blade element model}\label{sec:BE_model}

The REWS and REP models in Eqs.~\ref{eq:P_REWS} and \ref{eq:P_REP} consider the rotor to be a permeable actuator disc. In these models, airfoil characteristics and the rotation of the blades are abstracted with all turbine aerodynamic information contained in the coefficient of power $C_P$. This representation is inherently limited in its generalizability to handle nonlinear aerodynamic interactions between the wind turbine and wind shear. Blade element theory is an alternative approach that explicitly considers these factors in modeling turbine power production. The following is an overview of blade element theory and the process for evaluating the blade element (BE) model for an arbitrary set of wind speed and direction inputs.

\begin{figure*}
\centerline{\includegraphics[width=4.65in]{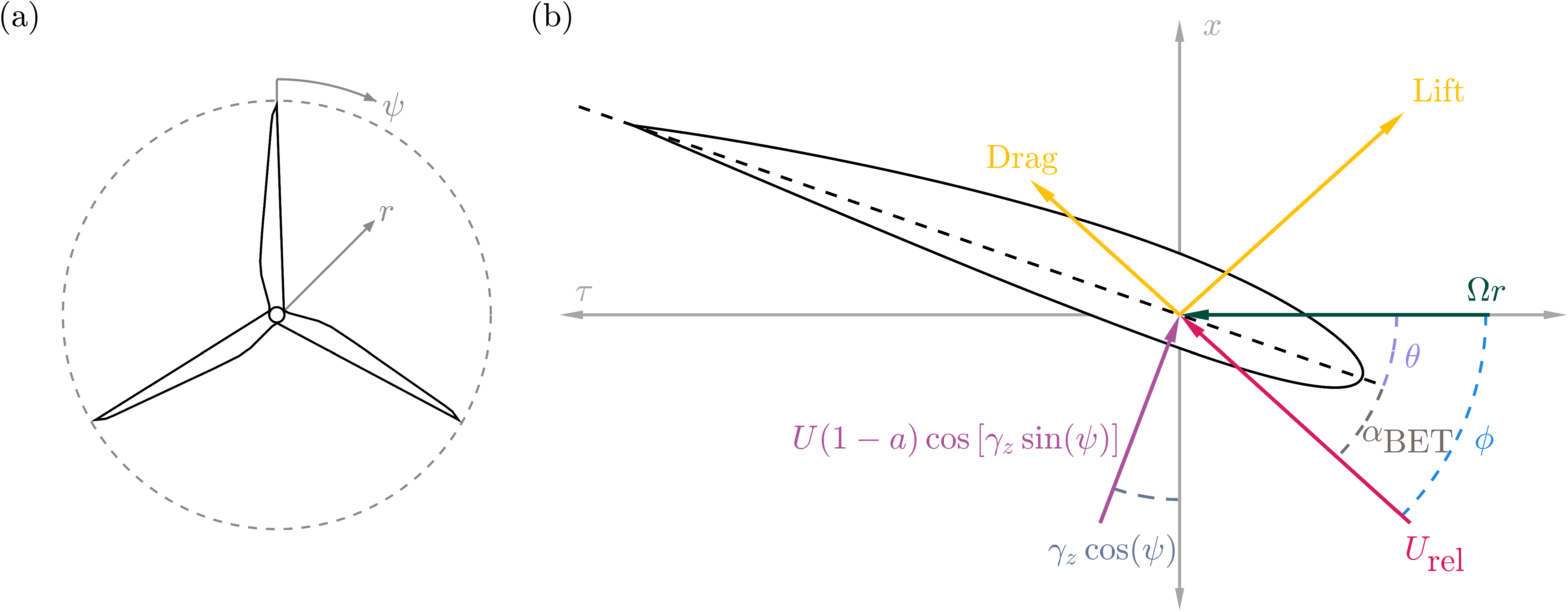}}
\caption{(a) Front view of a turbine rotor showing the conventions used for describing the position of the individual blade elements. The azimuth angle $\psi$ is measured from vertical and the local radial position $r$ from the axis of rotation. (b) A cross-sectional view of a blade element as it passes through azimuth angle $\psi = \ang{0}$. The axial and tangential directions are represented by $x$ and $\tau$, respectively. The relative wind $U_{\text{rel}}$ is computed from the axial and tangential components of the wind incident on the blade element. The axial component of the wind speed is a function of the freestream wind speed $U$ modified by the axial induction $a$, the local wind misalignment angle $\gamma_z$, and the azimuth angle of the blade element $\psi$. The tangential component of the velocity is a function of the angular velocity $\Omega$ and the local radial position $r$ of the blade element. The lift and drag forces are a function of the angle of attack, defined as $\alpha_{\text{BET}} = \phi - \theta$, where $\phi$ is the inflow angle of the relative velocity $U_{\text{rel}}$ and $\theta$ is the local pitch angle.\label{fig:BE_Diagram}}
\end{figure*}

In a blade element model of a turbine rotor, the blades are divided into discrete nodes. Because the airfoil geometry of the blades is a function of the radial position along the blade, each node will have a corresponding set of airfoil properties that govern the aerodynamics of that node. The basis of blade element theory assumes 1) the forces induced on a given node may be found independently of the other elements along the blade, and 2) that those forces are collectively a function of the airfoil geometry and the incident wind conditions at each element.~\cite{Branlard_Ch7,Weick_1930} As the blades rotate about the center axis with inflow containing wind speed and direction variations, the wind vectors incident on each blade node are a function of the azimuthal angle $\psi$ of each blade element, as well as the radial position $r$ of each element. Kragh and Hansen \cite{Kragh_2014} introduced a method for modeling the axial forces induced on each blade node accounting for speed shear and yaw misalignment. Howland et al.~\cite{Howland_2020} further generalized the model to account for speed shear, direction shear, and yaw misalignment in computing the axial and tangential forces acting on each blade element, along with a variable angular velocity that depends on the aerodynamic torque. This study uses the contributions of these studies in incorporating speed and direction shear into the blade element model.

What follows is a description of the sequence of calculations performed in evaluating the blade element model. The process described is repeated for each element along the turbine blade. The first step in evaluating the blade element model is to compute the axial and tangential components of the velocity incident on each node. Fig.~\ref{fig:BE_Diagram}(a) provides a depiction of the turbine rotor and the coordinate system used in evaluating the blade element model. Fig.~\ref{fig:BE_Diagram}(b) shows the incident wind vectors on a blade node as it passes through the vertical position ($\psi = \ang{0}$). The axial component of the velocity, given by

\begin{equation}\label{eq:U_axi}
U_{x}(r,\psi) = U(r,\psi) (1-a) \text{cos}\left[\gamma_z(r,\psi) \text{sin}(\psi) \right] \text{cos} \left[\gamma_z(r,\psi) \text{cos}(\psi) \right],
\end{equation}

\noindent is a function of the freestream velocity $U$ modified by the axial induction $a$, the local wind misalignment angle $\gamma_z$, and the azimuth angle of the blade element $\psi$. Axial induction depends on the specific induction closure model used, which we discuss in Sec.~\ref{sec:Induction_Controller}. The tangential component of the velocity is given by

\begin{equation}\label{eq:U_tan}
U_{\tau}(r,\psi) = \Omega r - U(r,\psi) (1-a) \text{cos} \left[\gamma_z(r,\psi) \text{sin}(\psi) \right] \text{sin} \left[\gamma_z(r,\psi) \text{cos}(\psi)\right],
\end{equation}

\noindent where $\Omega$ is the angular velocity of the blade element. Tangential induction is neglected (see Kragh and Hansen\cite{Kragh_2014}). For clarity of presentation, we hereafter omit the functional dependence of $U$ and other quantities derived from it on $r$ and $\psi$. The relative velocity incident on the blade element shown in Fig.~\ref{fig:BE_Diagram}(b) is a function of both the axial and tangential components of the velocity and is defined by the relationship

\begin{equation}\label{eq:U_rel}
U_{\text{rel}}^2 = U_x^2 + U_{\tau}^2.
\end{equation}

\noindent The inflow angle of the relative velocity vector in relation to the blade element is given by

\begin{equation}\label{eq:Phi}
\phi = \text{arctan} \left[\frac{U_x}{U_{\tau}}\right],
\end{equation}

\noindent and is used to compute the angle of attack on the airfoil, defined by

\begin{equation}\label{eq:AoA}
\alpha_{\text{BET}} = \phi - \theta,
\end{equation}

\noindent where $\theta$ is the pitch angle of the blade element. From the angle of attack $\alpha_{\text{BET}}$ (here stylized with ``BET'' to differentiate it from the speed shear exponent), the lift and drag coefficients ($C_L$ and $C_D$, respectively) are determined for the blade element. The lift and drag coefficients are typically determined empirically for a given airfoil geometry and tabulated as a function of the angle of attack. This step highlights a potential weakness in implementing the blade element model. Whereas the REWS and REP models require no information about the specific aerodynamic properties of the airfoils used on a given turbine, the BE model must necessarily have this information, which may be less readily available than $C_P$.

The tangential force induced by the incident wind conditions on a given blade element is defined by

\begin{equation}\label{eq:df}
\text{d}f_{\tau} = \frac{1}{2} B \rho U_{\text{rel}}^2 \left[ C_L(\alpha_{\text{BET}}) \text{sin} (\phi) - C_D(\alpha_{\text{BET}}) \text{cos} (\phi)\right]c\ \text{d}r,
\end{equation}

\noindent where $B$ is the number of blades and $c$ is the local chord length at each node. The incremental torque induced on each section by the tangential force is given by

\begin{equation}\label{eq:dQ}
\text{d}Q = r\ \text{d}f_{\tau},
\end{equation}

\noindent and the incremental power is then a function of the incremental torque and the rotor angular velocity, given by

\begin{equation}\label{eq:dP}
\text{d}P = \Omega \ \text{d}Q.
\end{equation}

\noindent Combining Eqs.~\ref{eq:df}-\ref{eq:dP} and integrating over the rotor area through one full rotation gives the combined rotor power, expressed as

\begin{equation}\label{eq:P_BE}
P_{\text{BE}} = \frac{1}{4 \pi} B \rho \Omega \int_0^{2\pi} \int_{0}^{R}  U^2_{\text{rel}} \left[ C_L(\alpha_{\text{BET}}) \text{sin} (\phi) - C_D(\alpha_{\text{BET}}) \text{cos} (\phi)\right]cr \ \text{d}r \ \text{d}\psi,
\end{equation}

\noindent which is the final expression for turbine power for a given arbitrary set of inflow conditions in the blade element model. The influence of wind speed and direction variations on the blade element model power predictions is distributed across several of the inputs listed above. The model first accounts for shear in the calculation of the relative velocity and accompanying inflow angle, which are functions of the axial and tangential components of the wind incident on each blade node. The inflow angle is then used to determine the angle of attack, which is subsequently used to determine the lift and drag coefficients at each node, which appear explicitly in Eq.~\ref{eq:P_BE}. By this mechanism, the effect of shear on the blade element model predictions cascades through the intermediate variables of the model, affecting the calculation of each subsequent variable appearing explicitly in the final equation for power in Eq.~\ref{eq:P_BE}.

\subsubsection{Induction closure and turbine controller modeling}\label{sec:Induction_Controller}

The BE model requires that the axial induction on the rotor $a$ and the rotor angular velocity $\Omega$ be specified, two requirements not present in the REWS and REP models shown above. In standard blade element momentum theory, the blade element equations shown in Sec.~\ref{sec:BE_model} are coupled with induction closure based on one-dimensional (streamwise) momentum theory: $a=(1-\sqrt{1-C_T})/2$, where $C_T$ is the thrust coefficient. However, since this relationship is derived under the assumption of uniform streamwise inflow (i.e., no wind speed or direction shear), it is unclear to what extent it is valid in environments that include wind shear. Further, the rotor angular velocity is not known a priori, and it may depend on the wind speed and direction shear. In this study, we consider two candidate closures for the axial induction and three possibilities for supplying the rotor angular velocity.

For induction closure, we consider modeling frameworks where: 1) axial induction is set as a constant $a = 1/3$ at the Betz limit value over the entire rotor (as in Kragh and Hansen\cite{Kragh_2014} and Howland et al.~\cite{Howland_2020}) and 2) the axial induction is closed locally over the rotor, depending on radial and azimuthal position, based on one-dimensional momentum theory. For angular velocity, we consider: 1) the angular velocity is set to maintain a constant tip-speed ratio, 2) the angular velocity is estimated by a controller model, and 3) the angular velocity is taken from the SCADA data. The reference BE case is the one of greatest simplicity, with the axial induction $a = 1/3$, and $\Omega$ set to the SCADA-measured values. In the results (Sec.~\ref{sec:Results}), for clarity of presentation, only the simple, reference BE model results are shown, and the influence of the different induction and rotor angular velocity modeling choices is shown separately in Sec.~\ref{sec:Induction_TSR_Analysis}. 

The localized induction model is implemented as described by Madsen et al.~\cite{Madsen_2020} In cases where the thrust coefficient exceeds a critical value of 1, momentum theory breaks down due to an inability to account for the effect of flow separation in the far wake that can cause velocities in that region to be negative.~\cite{Burton_Ch3,Branlard_Ch10} To address this, several empirical corrections have been proposed (see, e.g., Burton et al.~\cite{Burton_Ed2_Ch3}, Buhl\cite{Buhl_2005}, and Madsen et al.~\cite{Madsen_2020}). From Madsen et al.~\cite{Madsen_2020}, a single polynomial describing the axial induction as a function of the thrust coefficient is used. The coefficients of the polynomial are chosen to best fit standard one-dimensional momentum theory at low thrust coefficients, and to extrapolate to empirical results in high-thrust regimes. The localized induction model is given by

\begin{equation}\label{eq:CT_a_loc}
a(r,\psi) = 0.0883C_T^3(r,\psi) + 0.0586C_T^2(r,\psi) + 0.2460C_T(r,\psi),
\end{equation}

\noindent where the axial induction is now a function of radial and azimuthal position. The local thrust coefficient is defined as

\begin{equation}\label{eq:CT_loc}
C_T(r,\psi) = \cfrac{U^2_{\text{rel}} B c}{U(z_h)^2 2 \pi r} \left[ C_L(\alpha_{\text{BET}}) \text{cos} (\phi) + C_D(\alpha_{\text{BET}}) \text{sin} (\phi)\right].
\end{equation}

\noindent Tip and hub loss corrections are neglected for simplicity. A sensitivity analysis (not shown for brevity) indicated that inclusion of tip and hub loss corrections affected the results on the order of $<1$\%.

In the absence of SCADA data containing the turbine's realized tip-speed ratio, the BE model requires this information to be estimated. A common controller strategy in variable speed, pitch-regulated turbines modulates the generator torque to allow the rotor speed to vary proportionally to the inflow wind speed\cite{Burton_Ed2_Ch3} (referred to here as the $k\text{--}\Omega^2$ model). In Region~II, this model seeks to maintain a constant tip-speed ratio that theoretically maximizes power production in uniform inflow conditions. Because the turbine is tracking a fixed tip-speed ratio, the rotor speed $\Omega$ varies proportionally to the incoming wind speed. This in turn means that the torque on the rotor varies proportionally to $\Omega^2$. The aerodynamic torque is given by

\begin{equation}\label{eq:Q_aero}
Q_a = \frac{1}{2}\rho \pi R^5 \frac{C_P}{\lambda^3} \Omega^2.
\end{equation}

\noindent In steady-state operation, the controller attempts to keep the generator load torque equal to the aerodynamic torque by modulating the rotor speed. All variables on the left-hand side preceding $\Omega^2$ are constant, and often referred to collectively as $k$, hence the name of this model. For implementation of this controller model, the value of $k$ for this particular utility-scale turbine is computed by taking the slope of the line of best fit for the rotor torque as a function of the rotor angular velocity squared in the SCADA measurements. The resulting unique value of $k$ is used to compute the rotor speed, given by

\begin{equation}\label{eq:Omega_model}
k \Omega^2 = \frac{P_{\text{BE}}}{\Omega},
\end{equation}

\noindent for all input profiles with arbitrary wind speed and direction shear, where $P_{BE}$ is the power production estimated by the blade element model given in Eq.~\ref{eq:P_BE}.

The following is a summary of the sensitivity analysis performed here. For induction closure, we consider two cases: 1) constant induction ($a = 1/3$), and 2) the momentum theory closure described in Eq.~\ref{eq:CT_a_loc}. For the controller model, we consider three cases: 1) the angular velocity that maintains the target tip-speed ratio within the subset of region II determined from the SCADA data (referred to hereafter as $\lambda^*$), 2) the $k\text{--}\Omega^2$ model described above, and 3) the measured angular velocity from the SCADA data. Note that because of model discrepancy, using the estimated value of $k$ from the SCADA data does not necessarily achieve the tip-speed ratio that is targeted by the utility-scale turbine in the field setting ($\lambda^*$). We perform the sensitivity analysis with every combination of induction closure and turbine control. This scheme is designed to illustrate the influence of each model choice on the final BE model power predictions. Table \ref{Table:sensitivity} summarizes the six cases considered.

\begin{center}
\begin{table}[b]%
\centering
\caption{Summary of BE model sensitivity analysis setup.\label{Table:sensitivity}}%
\begin{tabular*}{250pt}{@{\extracolsep\fill}lcc@{\extracolsep\fill}}%
\toprule
\textbf{Case} & \textbf{Induction Closure} & \textbf{Controller Model} \\
\midrule
1 & Constant; $a = 1/3$ & $\lambda_{\text{SCADA}}$ \\
2 & Constant; $a = 1/3$ & $k\text{--}\Omega^2$ \\
3 & Constant; $a = 1/3$ & Constant; $\lambda = \lambda^*$ \\
4 & Momentum Theory     & $\lambda_{\text{SCADA}}$ \\
5 & Momentum Theory     & $k\text{--}\Omega^2$ \\
6 & Momentum Theory     & Constant; $\lambda = \lambda^*$ \\
\bottomrule
\end{tabular*}
\item A description of the induction closure and controller model used in each of the six cases tested in the sensitivity analysis on the BE model.
\end{table}
\end{center}

\subsection{Numerical implementation}\label{sec:Numerical_Implementation}

\begin{figure*}
\centerline{\includegraphics[width=\textwidth]{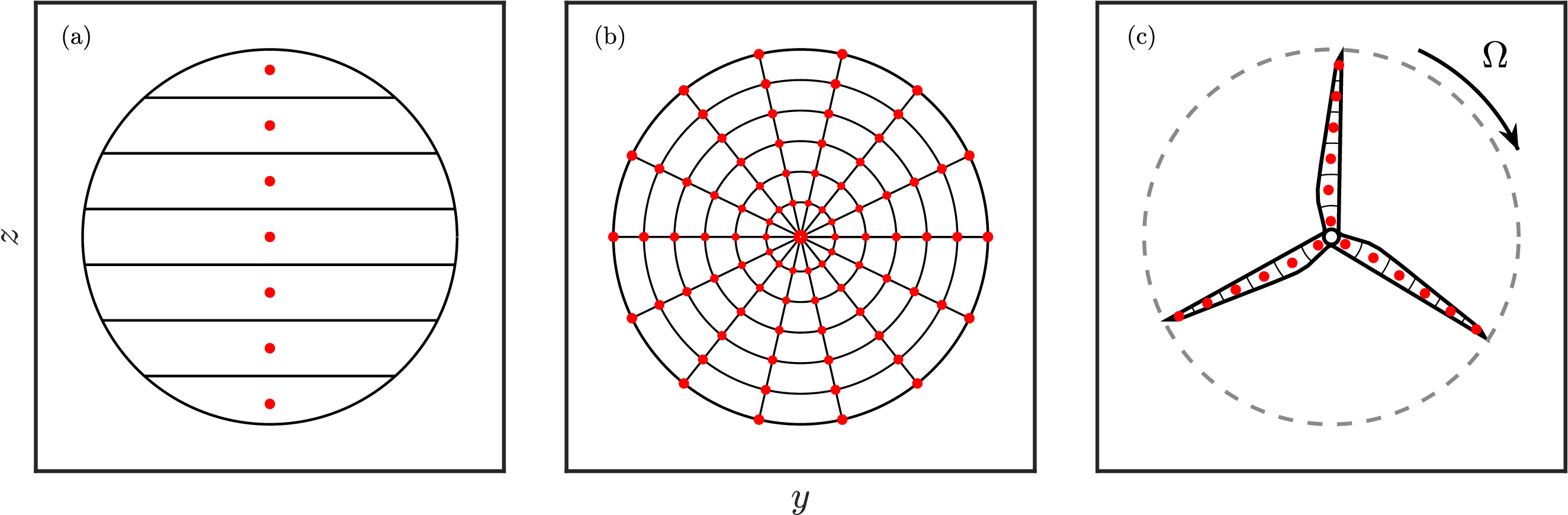}}
\caption{(a) Cartesian discretization of rotor disc area as described by Wagner et al.~\cite{Wagner_2009} (b) Polar discretization used in this study for the rotor-equivalent wind speed and rotor-equivalent power models. (c) Blade element representation of turbine rotor, showing the individual blade nodes at which the forces imparted by the wind are computed. The number of blade nodes depicted is reduced for visual clarity. Red markers in each subplot indicate the locations at which the wind field is evaluated.\label{fig:Disc}}
\end{figure*}

Fig.~\ref{fig:Disc} shows illustrations of the rotor discretization schemes considered in this study. Fig.~\ref{fig:Disc}(a) shows the method described by Wagner et al.~\cite{Wagner_2009} in which the rotor is divided into a series of horizontal segments of varying area in which individual measurements of the wind field are taken. The measurements of the wind field in each horizontal segment are assumed to be constant over the segment area. The resulting rotor-equivalent wind speed is then computed as the weighted average where each measurement of wind speed is weighted by the fraction of the total rotor area that its respective segment constitutes. Fig.~\ref{fig:Disc}(b) shows the discretization method used for the two rotor-equivalent models in this study. We discretize the rotor into a series of concentric circles emanating from the center and draw a series of radii extending out at evenly spaced azimuth angles. The wind field is evaluated at the intersections of these annuli and radii. Fig.~\ref{fig:Disc}(c) shows a representation of the blade element model for the turbine rotor. As noted in Sec.~\ref{sec:BE_model}, blade element theory models the blades of the rotor and considers that they are moving through space in time, and therefore experience different degrees of wind shear as they transit a rotation of the rotor. The red markers in each subplot of Fig.~\ref{fig:Disc} indicate the locations at which the wind field is evaluated for each rotor discretization scheme. The models presented in Sec.~\ref{sec:turbine_models} assume tilt and cone angles are zero.

The wind turbine models can be applied with instantaneous values of wind speed $U(r,\psi)$ and direction $\Lambda(r,\psi)$. However, given the inherent challenges of measuring two-dimensional instantaneous wind fields in full-scale field environments, these models are often driven by finite time-averaged wind measurements.~\cite{Howland_2020,howland2022collective,Wagner_2009,Choukulkar_2016} To account for kinetic energy contributions of turbulence in 10 min averaged wind measurements, Wagner et al.~\cite{Wagner_2009} and Choukulkar et al.~\cite{Choukulkar_2016} included turbulence intensity in their derivations of the REWS and REP models, with the turbulence intensity characterizing turbulence energetics at timescales smaller than the 10 min time-averaging scale. However, as noted by Choukulkar et al.~\cite{Choukulkar_2016}, the influence of turbulence on the rotor aerodynamics (e.g., $C_P$, $C_L$,$ C_D$) is neglected.~\cite{akhmatov2007influence} To be consistent with the shear-dependent REWS, REP, and BE formulations in the previous subsections, turbulence must be known at each position in the rotor $(r,\psi)$. In the present field measurements, since power depends non-linearly on the instantaneous flow, we use 1 min averaged wind measurements rather than 10 min measurements.~\cite{howland2022collective} Further, turbulence intensity at timescales below 1 min averaging is not available from the profiling LiDAR, but turbulence intensity is estimated at hub-height ($z=z_h$) in the wind turbine SCADA data at 1 min resolution. Therefore, the models could be supplemented with the SCADA turbulence intensity, assuming a constant value over the rotor area. The effect of including the hub height turbulence intensity on the root mean square error (RMSE) of the model predictions in this study is on the order of 1\%. Since this does not change the qualitative or quantitative conclusions of this study, we therefore neglect turbulence intensity in the model formulations to focus on the primary effects of wind speed and direction shear. We also neglect consideration of turbulence intensity to maintain a controlled experiment between the two rotor-equivalent models and the blade element model, which does not presently incorporate information about turbulence. Future work should revisit this analysis with measurements of wind speed, direction, and turbulence kinetic energy  across the rotor area in $(r,\psi)$. Hereafter, we refer to 1 min averaged wind speed and direction measurements as $U(r,\psi)$ and $\Lambda(r,\psi)$, respectively.

\section{Wind farm setup and experimental design}\label{sec:Exp_Design}

The wind farm in this study is located in northwestern India. The terrain at the site is primarily flat and changes in elevation by approximately 100 m over the extent of the farm, which is a distance of about 25 km. The farm contains approximately 100 utility-scale horizontal axis wind turbines from a variety of manufacturers. The hub height and rotor diameter of each turbine are both approximately 100 m.~\cite{Howland_2020}

Measurements of wind velocity at the site were taken with a Vaisala Leosphere Windcube V2.0 aerosol backscatter Doppler LiDAR system. This device determines the speed and direction of the wind at predetermined elevations by measuring the Doppler shift in infrared light reflected off of particles in the atmosphere from successive pulses.~\cite{lindeloew_2009} Concurrent with the LiDAR measurements, power production, nacelle heading, blade pitch angle, rotor angular velocity, and other operational parameters for each turbine in the array are recorded by the onboard SCADA system. The experiment was conducted from February to April 2020. LiDAR wind profile data and SCADA turbine data are collated in 1 min averages.

The empirical results in this study use the SCADA data reported by the first turbine downstream from the LiDAR, located approximately 250 m, about 2 rotor diameters, away.~\cite{Howland_2020} For the Region~II wind speeds considered in this study, the advection timescale between the LiDAR and the turbine of interest is on the order of 20--30 seconds. Given the topography of the farm, we assume that the atmospheric conditions are horizontally homogeneous. While the atmosphere is seldom truly homogeneous in reality, this is a common model for the ABL provided the spatial domain is sufficiently large and flat as is the case at the site (see discussion in Stull \cite{Stull_ch3} and Wyngaard \cite{Wyngaard_ch5}). While the turbine commands zero yaw misalignment, due to the time lag in the yaw controller, this does not guarantee a zero-degree yaw offset. The local wind misalignment angle defined in Eq.~\ref{eq:gamma} accounts for this by incorporating the turbine-measured nacelle heading rather than assuming zero yaw offset. Wind profiles in which the hub height wind speed measured by LiDAR was outside of the inner section of region II (6--8 $\text{ m s}^{-1}$) were also excluded.

\subsection{Site characterization}\label{sec:Site_Characterization}

To determine the effect of discrete combinations of speed and direction shear on turbine power production, we impose a standardized procedure to classify the degree of each type of shear observed in the LiDAR wind profile measurements. In meteorological and wind energy applications, two methods are commonly used to model wind speed as a function of height in the ABL. The first of these, the log wind profile, is a semi-empirical relationship derived from similarity theory in the surface layer, which models the wind speed as a function of the surface roughness length and surface friction velocity at a given location (see additional discussion in Lundquist\cite{Lundquist_2020} and Peinke et al.~\cite{Peinke_Ch11}). Since we focus on wind shear, at least in part, above the surface layer in this study, we use the power law, defined as

\begin{equation}\label{eq:Speed_Shear}
U(z) = U_{\text{ref}} \left[ \frac{z}{z_{\text{ref}}} \right]^{\alpha},
\end{equation}

\noindent where the wind shear exponent $\alpha$ is fit to each wind speed profile using least-squares regression. The reference height $z_{\text{ref}}$ is the lowest elevation of the rotor and the reference wind speed $U_{\text{ref}}$ is the corresponding wind speed at that height.~\cite{Lundquist_2020,MSG_2020} The power law exponent $\alpha$ therefore serves as a non-dimensional quantification of the degree of speed shear in each profile. The degree of direction shear over the rotor is characterized by the average degree of turning from the bottom to the top of the rotor, given by 

\begin{equation}\label{eq:Dir_Shear}
\beta = \frac{\gamma_z(z = z_h + R) - \gamma_z(z = z_h - R)}{2R}.
\end{equation}

\noindent By convention, negative direction shear ($\beta < \ang{0}\text{m}^{-1}$) indicates counterclockwise turning over the rotor, usually referred to as backing, and positive shear ($\beta > \ang{0}\text{m}^{-1}$) indicates clockwise turning, termed veering.~\cite{Lundquist_2020} For Eq.~\ref{eq:Dir_Shear} to be an accurate description of the wind direction variation over the rotor, the wind direction gradient must be approximately linear. To justify the assumption of linear direction shear in this study, we note 1) a review of the relevant literature for modeling the evolution of wind direction with height yields no generally accepted model (as exists for speed shear), and 2) our analysis shows that for our site, the assumption of linear direction shear is, in aggregate, more accurate than assuming no shear. In this analysis, we compute the RMSE between the LiDAR wind direction profiles and 1) profiles that contain no direction shear over the rotor area, and 2) profiles that vary linearly between the wind direction at planes located at the bottom and top of the rotor. The distribution of RMSE for the model assuming no shear shows greater positive skew and a larger median RMSE ($\ang{3.28}$) when compared to the model assuming linear shear ($\ang{1.51}$). We use the above characterization of shear to qualitatively assess how discrete combinations of speed and direction shear affect power production observed in the empirical field measurements and in the model predictions.

\begin{figure*}
\centerline{\includegraphics[width=\textwidth]{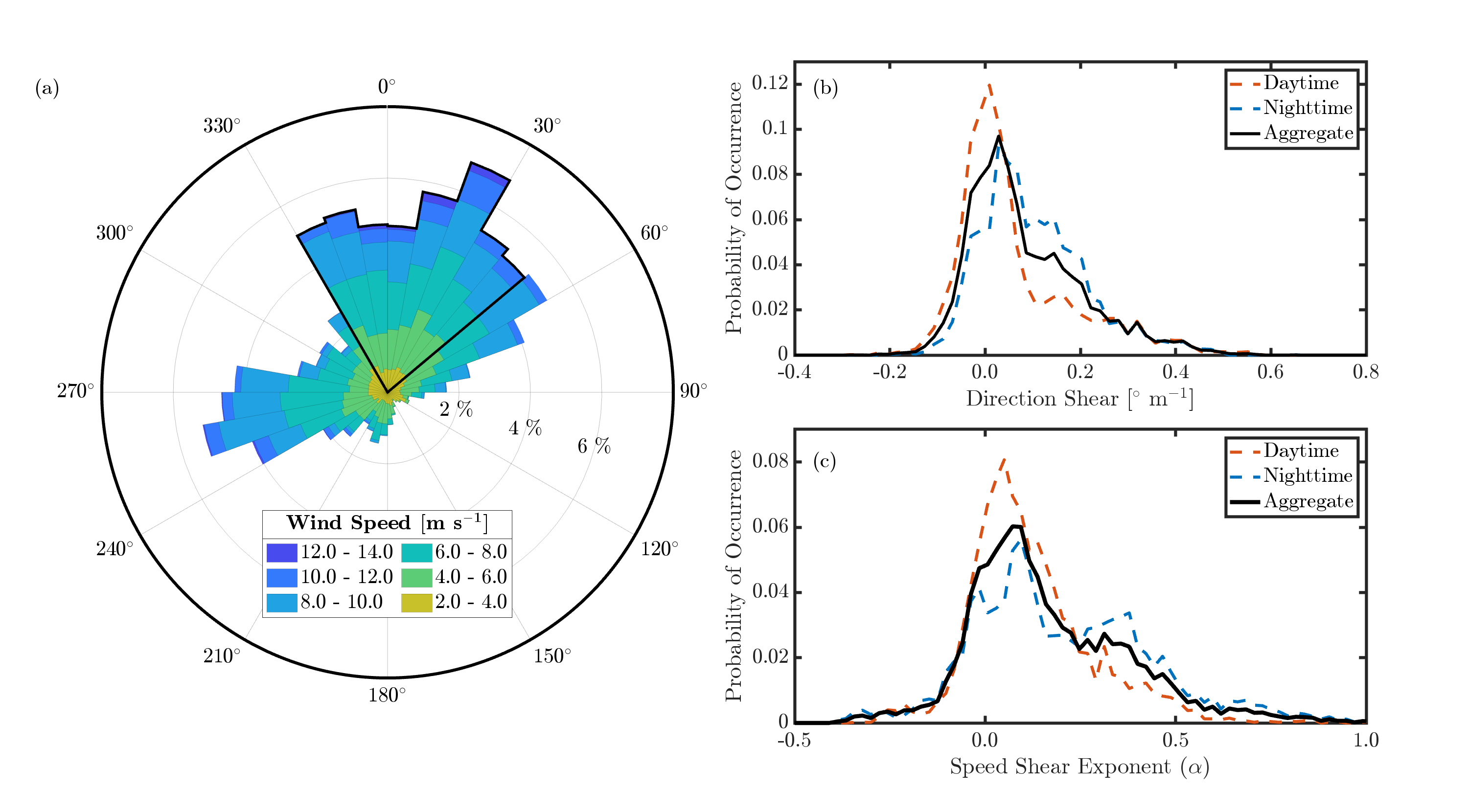}}
\caption{(a) Wind rose showing distribution of hub height wind speed and direction at the site. Northerly winds inside the black border are retained for subsequent analysis, those outside are excluded to control for potential wake effects of nearby turbines. Probability distribution of the degree of (b) direction shear and (c) speed shear over the turbine rotor area. Daytime and nighttime are defined by sunrise and sunset at the site during the period of the experiment.\label{fig:Site_Characterization}}
\end{figure*}

Wind conditions at the site are dominated by northerly and westerly winds during the period of the experiment (see Fig.~\ref{fig:Site_Characterization}(a)). Only northerly winds ($330^{\circ}$ to $45^{\circ}$) were retained for analysis to eliminate instances where wake effects from nearby turbines may be present in the inflow wind conditions. Probability distributions of speed and direction shear (see Fig.~\ref{fig:Site_Characterization}(b) and \ref{fig:Site_Characterization}(c)) show strong diurnal variations. Speed shear tends to be lower during the daytime hours (from sunrise to sunset), consistent with the expectation that turning is generally low in the presence of a convective boundary layer. For both daytime and nighttime, the frequency of positive speed shear is greater than that of negative speed shear. The nighttime speed shear distribution is bimodal, indicating a prevalence of shear around $\alpha = 0.2$ and $\alpha = 0.4$. We also note the presence of low-level jets (LLJs) at the site. A LLJ may occur in the stratified flows occurring at night when inertial oscillations generated by the release of turbulent stresses from the decay of the convective boundary layer align with the mean wind to form a jet near to the ground.~\cite{Lundquist_2020} An analysis following that described by  Debnath et al.~\cite{Debnath_2023} for systematically identifying the presence of low-level jets reveals that LLJs are present in 8\% of the 1 min time-averaged wind profile measurements. This analysis applied the following conditions for determining that a LLJ was present: 1) the wind speed maximum is inside the rotor area, 2) the absolute dropoff from the wind speed maximum to the local minimum above the core is $>2 \text{ m s}^{-1}$, and 3) and the ratio of the dropoff to the wind speed maximum is $>10$\%. The combination of both dimensional and non-dimensional conditions is supported by previous studies (see, e.g., Aird et al.~\cite{Aird_2021}, Kalverla et al.~\cite{Kalverla_2019}, and Luiz and Fiedler\cite{Luiz_2022}). The complex shear profiles that can occur in the rotor area as a result of an LLJ may affect the accuracy of the power law fit in several ways. Especially large magnitudes of shear in parts of the rotor may bias the shear exponent which is based on a least-squares fit across the entire profile, indicating a higher or lower degree of shear than is present in different portions of the wind profile. Conversely, positive and negative shear occurring simultaneously over the rotor area may effectively cancel out in a least-squares fit constrained to the form of the power law profile equation. This causes the shear exponent to over-estimate shear in the negative shear region of the rotor area while simultaneously under-estimating shear in the positive shear region. This highlights a crucial deficiency with the characterization of speed shear with the power law. More generally, it is difficult to characterize wind profiles that take on many complex shapes with one or two parameter wind profile models. This again motivates the development of an accurate parametric wind power model that can take arbitrary wind fields as input. Therefore, site and time-varying wind profiles can be provided to the power model to assess their impact on power and energy production more broadly without requiring qualitative characterization.

Daytime direction shear is most often around $\beta = \ang{0.1}\text{m}^{-1}$, with similar degrees of backing and veering overall. Nighttime direction shear is generally larger and more often veering than backing. Again, this could be indicative of Ekman-type flows associated with LLJs in the stable boundary layer, although we note it is not a precondition that an LLJ be present for direction shear to occur.~\cite{Stull_ch6}

\subsection{Wind profile model inputs}\label{sec:Model_Inputs}

\begin{figure*}
\centerline{\includegraphics[width=4.5in]{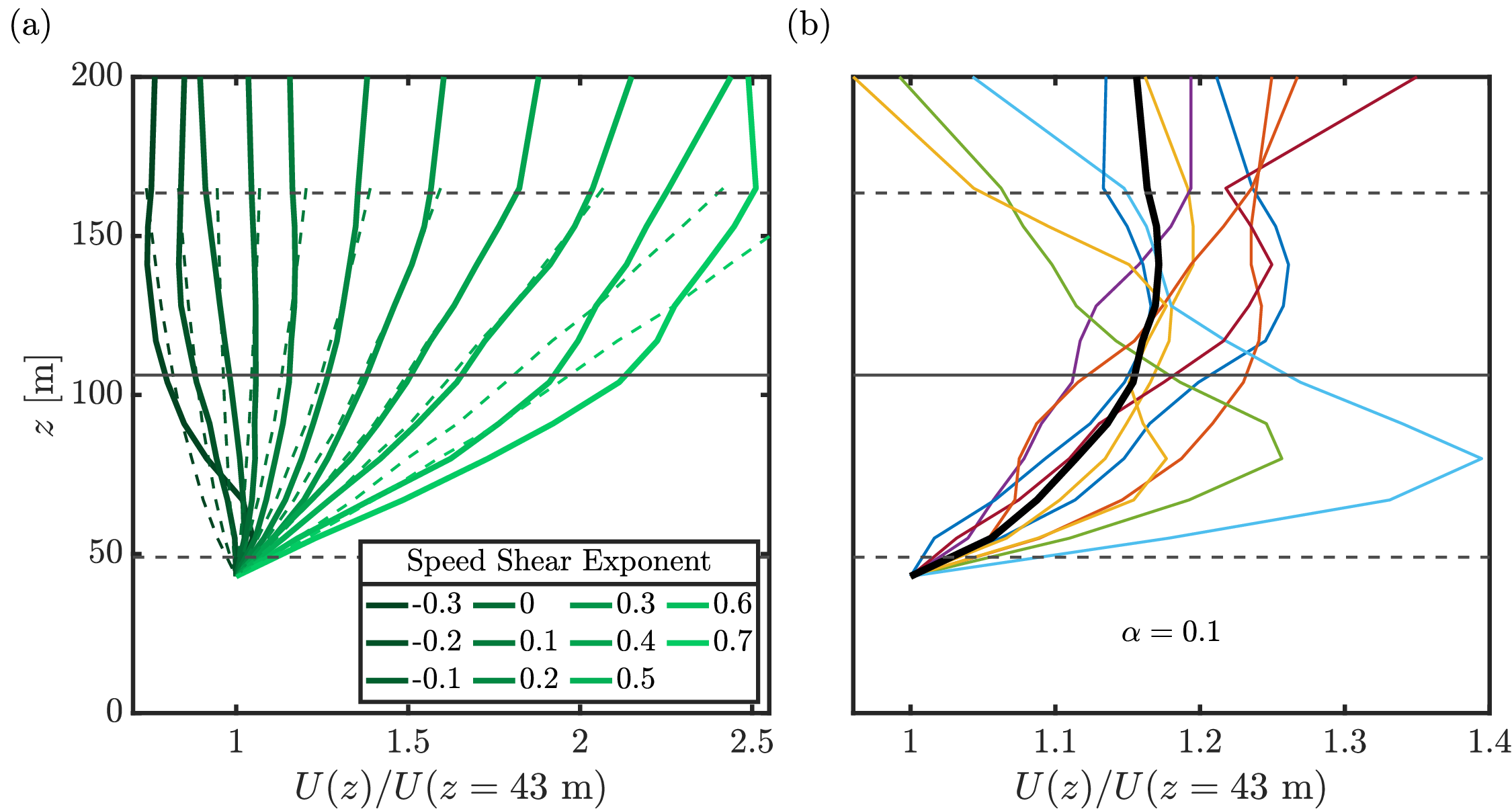}}
\caption{Panel (a) shows median wind speed profiles (solid lines) measured by LiDAR binned by shear exponent value and the corresponding power law profiles (dashed lines) fit to each median profile. Panel (b) shows the median wind speed profile in black and 10 randomly selected 1 min time-averaged wind speed profiles in color within the same best-fit wind speed shear exponent ($\alpha = 0.1$). These profiles demonstrate significant heterogeneity despite being characterized by the same speed shear exponent. The dashed grey lines indicate the top and bottom elevations of the turbine rotor. Note the horizontal scaling in (b) is modified to highlight variation in individual 1 min profiles.\label{fig:Speed_Profiles}}
\end{figure*}

We show in Fig.~\ref{fig:Speed_Profiles}(a) that while the power law can be an appropriate approximation of the wind speed profiles in a median sense, instantaneous wind profiles may exhibit the effect of turbulent (microscale) and mesoscale transient variations such as LLJs that cause the profiles to deviate from the power law. In Fig.~\ref{fig:Speed_Profiles}(b), the median wind speed profile associated with the wind speed shear $\alpha=0.1$ bin is shown, along with 10 randomly selected 1 min averaged wind speed profiles that are also characterized through best-fit power laws by $\alpha=0.1$. While a power law with $\alpha=0.1$ is descriptive of the median profile, many finite time-averaged profiles exhibit complex vertical trends, including LLJs, that are insufficiently captured by the power law with $\alpha=0.1$.

To investigate sensitivities in the power predictions produced by the models based on deviations in the wind speed and direction profiles from their canonical representations, we test each model with two separate classes of inputs. First, each model is tested with input canonical ABL profiles where wind speed as a function of height is represented exactly by the best-fit power law relationship and wind direction shear is perfectly linear over the rotor area (see Choukulkar et al.~\cite{Choukulkar_2016} and Walter\cite{Walter_2007}). Second, because these canonical models are not descriptive of the true profiles in the ABL for short time averages (as shown in Fig.~\ref{fig:Speed_Profiles}), each model is tested with the observed LiDAR wind speed and direction measurements from the profiling LiDAR. This numerical experiment will demonstrate the degree to which power predictions stemming from a power law approximation can represent the power produced by a utility-scale turbine.

\subsection{Power production normalization}\label{Sec:Normalization}

\begin{figure*}
\centerline{\includegraphics[width=3.53in]{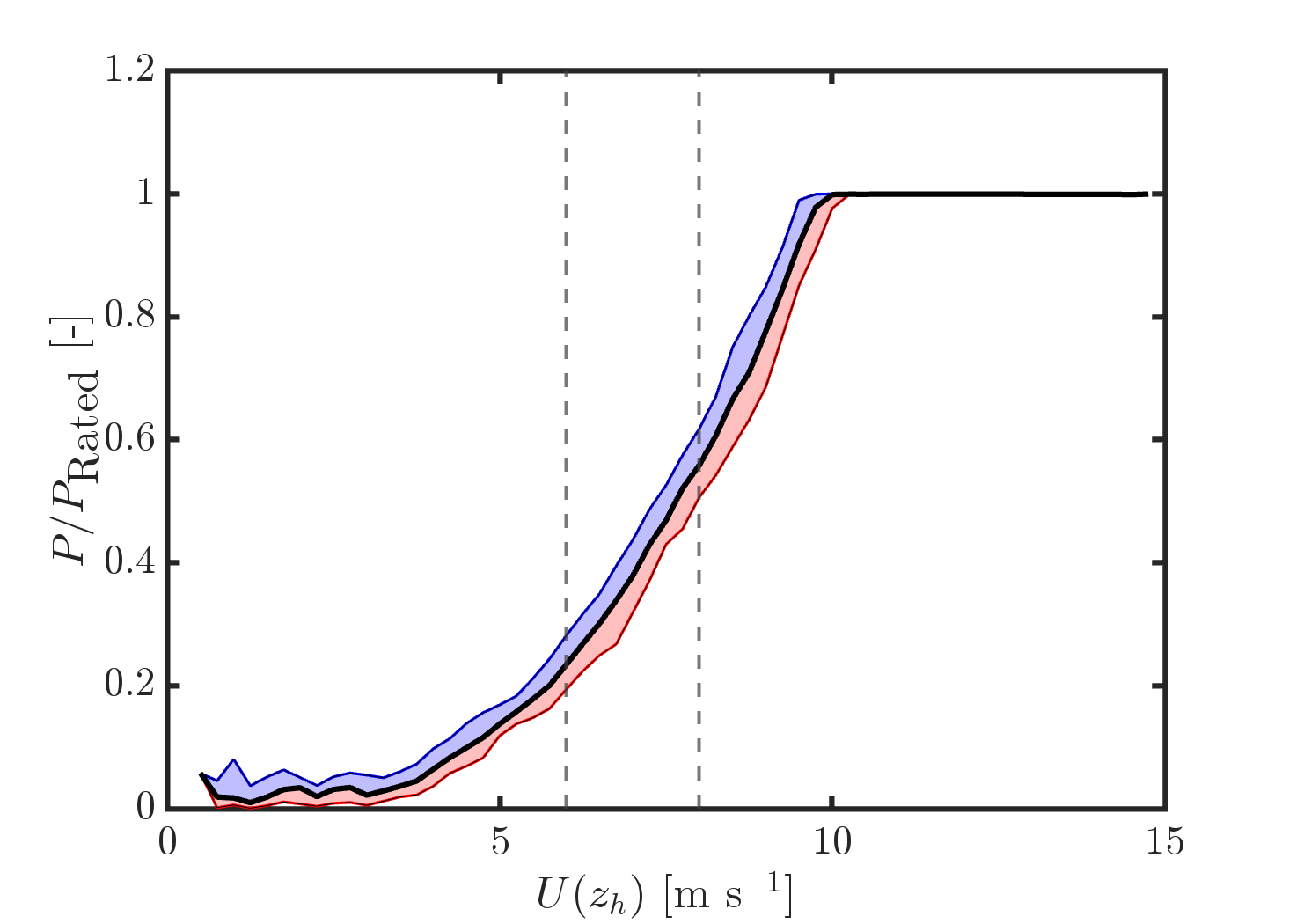}}
\caption{The median power curve $P^*$ for the turbine at the site. Measurements of power from SCADA data are binned by their corresponding hub height wind speed $U(z_h)$ measured by LiDAR in increments of $0.5 \text{ m s}^{-1}$. The bold black line is the median power curve, and the red and blue shaded regions indicate the 25\% and 75\% quantiles around the median. The dashed lines indicate the subset of region II from which the LiDAR wind profiles used in this study are taken.\label{fig:Median_PCurve}}
\end{figure*}

Eq.~\ref{eq:P_Aero} shows that power scales as the cube of the wind speed. This relationship tends to dominate the effect of other atmospheric determinants of power, including speed and direction shear. To isolate the effect of shear on turbine power production, values of power are normalized to account for the nonlinear relationship between power and wind speed. The basis for this normalization is the median power curve, which we construct from the SCADA power measurements and their associated LiDAR wind profile measurements. SCADA power measurements are binned based on the corresponding hub height wind speed measured by LiDAR in increments of $0.5 \text{ m s}^{-1}$ and the median is computed for each bin. The resulting curve, shown in Fig.~\ref{fig:Median_PCurve}, is the median power production of the turbine, denoted by $P^*$, and is a function of hub height wind speed $U(z_h)$ measured by the LiDAR. Normalized SCADA power measurements and normalized model power predictions from LiDAR wind profile inputs are given by

\begin{equation}\label{eq:P_1xNON}
\hat{P} = \frac{P}{P^*(U(z_h))},
\end{equation}

\noindent where $P$ is dimensional power and the denominator is the median power curve evaluated at the corresponding hub height wind speed for each input wind profile. Values of normalized power $\hat{P}$ relate each SCADA measurement or model prediction of power production to the median operation of the turbine across all wind conditions encountered. This procedure accounts for the cubic dependence of power on wind speed and the fact that two wind profiles with a different range of wind speed magnitudes over the rotor area may have the same degree of speed and direction shear as determined by Eqs.~\ref{eq:Speed_Shear} and \ref{eq:Dir_Shear}. Subsequent discussion of the results will reference over-performance with respect to the median power curve power production, which is defined to be values of normalized power greater than one ($\hat{P} > 1$), and under-performance, or values of normalized power less than one ($\hat{P} < 1$). 

Following the methodology proposed by Sanchez Gomez and Lundquist~\cite{MSG_2020}, to assess how discrete combinations of speed and direction shear affect power production, we group empirical power measurements based on the degree of shear in their corresponding wind profiles as determined by Eqs.~\ref{eq:Speed_Shear} and \ref{eq:Dir_Shear}. We define bins of width 0.1 for $\alpha$ and $\ang{0.1}\text{m}^{-1}$ for $\beta$. For SCADA measurements and model predictions from LiDAR wind profile inputs, we compute the mean value of all normalized measurements of power $\hat{P}$ placed into that bin based on the degree of speed and direction shear in the input wind profiles. The resulting bin value is given by

\begin{equation}\label{eq:P_Bin_hat}
\hat{P}_{\text{Bin}} = \overline{\hat{P}(\beta_i,\alpha_j)},
\end{equation}

where $\hat{P}_{\text{Bin}}$ is non-dimensional, the overbar indicates the mean across the independent 1-min averaged samples contained within the bin, and $i$ and $j$ are the indices for the bins for the wind direction shear and wind speed shear, respectively. 

Conversely, for results from the canonical ABL wind profile inputs, each bin includes a single power prediction generated by the models from the input wind profiles with the degree of speed and direction shear for each respective bin (i.e. there is only one value of the hub height wind speed set across all model predictions when driven by canonical power law profiles because the wind speed magnitude does not affect the normalized power response in Region~II). Thus, the bin value in this case is given by

\begin{equation}\label{eq:P_Bin_M}
P_{\text{Bin}}^{\text{M}} = P(\beta_i,\alpha_j),
\end{equation}

\noindent where the superscript ``M'' in $P_{\text{Bin}}^{\text{M}}$ denotes that each bin contains only a single dimensional power prediction from the models.

To assess the influence of complexities in the wind profiles not represented by the power law, we introduce a second normalization that enables a comparison between the power predictions from the simplified canonical ABL inputs with the empirical results. This second normalization addresses the fact that the bin values for empirical results ($\hat{P}_{\text{Bin}}$) have been non-dimensionalized by Eq.~\ref{eq:P_1xNON} while the model predictions from the canonical ABL inputs ($P_{\text{Bin}}^{\text{M}}$) have not. This test is intended foremost to identify the sensitivity of the model predictions to variations in the wind speed and direction profiles and deviations from their canonical ABL representations. By extension, this also demonstrates the degree of potential error that may occur in wind resource assessment performed using these models if canonical ABL profiles are assumed, provided empirically measured wind profiles are unavailable at a given site, or micro-siting location. The resulting normalized empirical results and power predictions from canonical ABL profile inputs are given by

\begin{equation}\label{eq:P_Bin_tilde}
  \tilde{P}_{\text{Bin}}(\beta_i,\alpha_j) =
  \begin{cases}
    \cfrac{\hat{P}_{\text{Bin}}(\beta_i,\alpha_j)}{\hat{P}_{\text{Bin}}(\beta = \ang{0}\text{m}^{-1},\alpha=0)} & \parbox[t]{0.40\textwidth}{
                    for empirical SCADA power measurements and power\\
                    predictions from LiDAR wind profile inputs,} \\[10pt]
    \cfrac{P_{\text{Bin}}^{\text{M}}(\beta_i,\alpha_j)}{P_{\text{Bin}}^{\text{M}}(\beta = \ang{0}\text{m}^{-1},\alpha=0)} & \text{for power predictions from canonical ABL profile inputs.} \\
  \end{cases}
\end{equation}

\noindent This normalization relates the power predictions for all combinations of shear to the case where the inflow contains no speed or direction shear. In all figures where the normalization described in Eq.~\ref{eq:P_Bin_tilde} has been applied, the no-shear power prediction (with a normalized value of $1$) has been removed to indicate that trends in power production are with respect to this value, not the median power curve. A more detailed discussion of the differences in the model predictions between the two inputs classes and in relation to the SCADA measurements is given in Sec.~\ref{sec:Discussion}.

\section{Results}\label{sec:Results}

In the following section, we discuss the results pertaining to the two primary goals in this study. First, we discuss the trends in power production observed in the SCADA measurements when the degree of speed and direction shear present in the corresponding LiDAR wind profile measurements is characterized according to Eqs.~\ref{eq:Speed_Shear} and \ref{eq:Dir_Shear}. We then move to a discussion of the model predictions. This includes a quantitative evaluation of the correlation and overall error of the models with respect to the SCADA measurements, as well as trends in model predictions for each of the two wind profile input classes. Presented first are the model predictions when driven with the best-fit canonical ABL profile inputs, followed by the model predictions produced by LiDAR wind profile inputs. A quantitative comparison of the model predictions across the two input classes is provided. We also relate the error for each of the three models that have knowledge of variations in speed and direction over the rotor area to the hub height wind speed model, which serves as a benchmark for the other models as it does not consider shear. We end by presenting the results of the sensitivity analysis on the BE model with respect to the choice of both the induction closure and turbine controller models.

\subsection{Empirical}\label{sec:Results_Empirical}

\begin{figure*}
\centerline{\includegraphics[width=\textwidth]{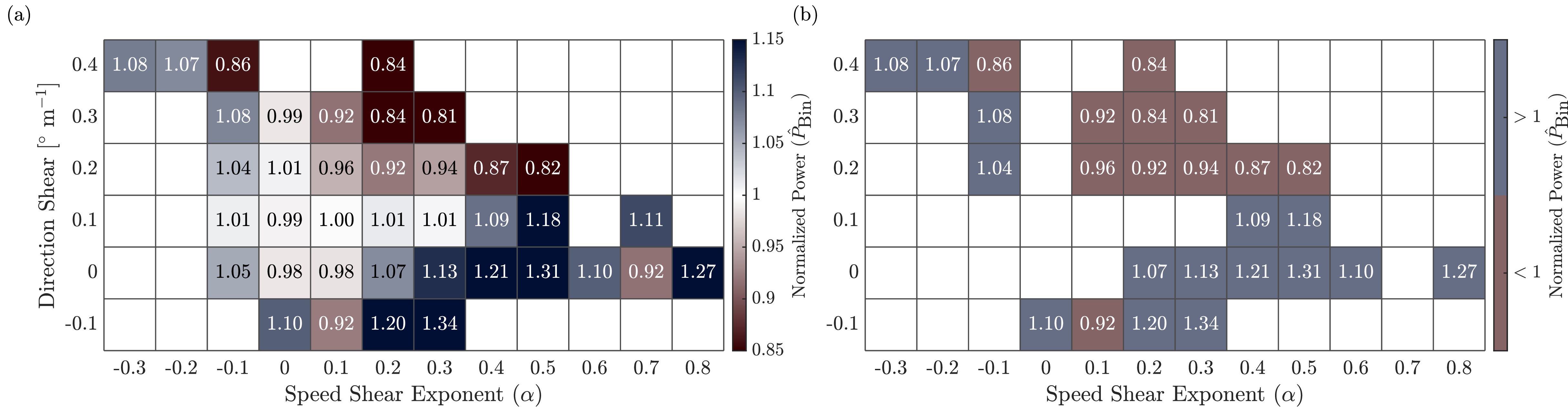}}
\caption{Bin averages of normalized power ($\hat{P}_{\text{Bin}}$) with respect to the median power curve of the turbine at the site for discrete combinations of speed and direction shear. Panel (a) shows empirical results for normalized power production from the turbine at the site. Only bins containing 30 or more power predictions are shown. Panel (b) shows the same data where the colormap is adjusted to highlight trends in over and under-performance. Conditional averages where the 95\% confidence interval around the mean, as estimated by bootstrapping, includes the value $\hat{P}_{\text{Bin}} = 1$ are removed.\label{fig:Empirical_Contour}}
\end{figure*}

Fig.~\ref{fig:Empirical_Contour} shows the effect of speed and direction shear observed in the SCADA power measurements described in Sec.~\ref{sec:Site_Characterization}. The values shown are the mean of all normalized power observations ($\hat{P}_{\text{Bin}}$) within each bin representing a specific combination of speed and direction shear. Only bins containing 30 or more data points are shown. Greatest over-performance to the median power curve is observed in the region bounded by $\alpha > 0.1$ and $\beta < \ang{0.2}\text{m}^{-1}$. As direction shear increases, the mean normalized power tends to decrease, with the lowest values observed in the upper right corner, corresponding to speed shear of about $\alpha = 0.3$ and direction shear of about $\beta = \ang{0.3}\text{m}^{-1}$. To demonstrate the statistical significance of the trends in over and under-performance shown in Fig.~\ref{fig:Empirical_Contour}(a), we compute the 95\% confidence interval around the mean values in each bin using bootstrapping. Fig.~\ref{fig:Empirical_Contour}(b) subsequently shows the empirical results where squares in which the confidence interval includes the value of normalized power $\hat{P}_{\text{Bin}} = 1$ are removed.

The results in Fig.~\ref{fig:Empirical_Contour} show qualitative similarities to those observed by Sanchez Gomez and Lundquist \cite{MSG_2020} in a study performed at a wind farm in central Iowa. The region of highest over-performance located in the region bounded by $\alpha > 0.1$ and $\beta < \ang{0.2}\text{m}^{-1}$ is characteristic of both sites. However, Sanchez Gomez and Lundquist did not observe wind shear falling into the region bounded by $\alpha < 0$ and $\beta < \ang{0.1}\text{m}^{-1}$, meaning that the over-performance observed in this region in Fig.~\ref{fig:Empirical_Contour}(a) cannot be directly compared across sites. Another notable difference in results between these two studies is the location of the greatest observed under-performance. Whereas the results in this study again show greatest under-performance in the region bounded by $\alpha > 0.1$ and $\beta > \ang{0.1}\text{m}^{-1}$, Sanchez Gomez and Lundquist observed greatest under-performance in the region bounded by $\alpha < 0.2$ and $\beta > \ang{0.1}\text{m}^{-1}$, similar to, but distinct from the results observed in this study. Several factors could influence these differences, including different rates of LLJs between the two sites, a factor which complicates analysis of these plots as noted in Sec.~\ref{sec:Site_Characterization}, and which will be discussed further in Sec.~\ref{sec:Discussion}.

Another key difference is the time averaging window for the LiDAR and SCADA data. This study uses 1 min time-averaged SCADA power measurements and LiDAR wind profile measurements, whereas the analysis by Sanchez Gomez and Lundquist\cite{MSG_2020} used 10 min time-averaged data. By averaging over a larger time window, the tails of the distributions for both speed and direction shear are reduced as wind profiles approach their respective median representations, as shown in Fig.~\ref{fig:Speed_Profiles}. Altering the time averaging window may affect the trends observed by again shifting data between bins, potentially altering the mean bin value displayed, consequently changing whether those values indicate over or under-performance. We therefore note that the conclusions that we draw from these plots are vulnerable to the method of shear characterization and the data processing methodology used in their generation. The utility this method presents is primarily in its ability to indicate qualitative trends in the effect of shear on power production across the range of speed and direction shear considered here, rather than predicting quantitative trends of over and under-performance resulting from an input wind profile with a given combination of speed and direction shear. Quantitative predictions depending on arbitrary speed and direction shear are the focus of the modeling in Sec.~\ref{sec:Results_Models}.

\subsection{Model predictions}\label{sec:Results_Models}

Given the complexity of the qualitative description of the wind profiles and of the empirical results, we investigate parametric models to predict the influence of wind shear on power production. In assessing the models, we wish to address three questions: (1) how do predictions of power vary based on the two input types used to drive the models, (2) how accurately do the models predict power based on arbitrary inflow conditions containing wind speed and direction shear from the LiDAR measurements, and (3) how accurately do the models predict power when the effect of shear is isolated from the dominant effect of the wind speed magnitude? To answer the first question, we qualitatively and quantitatively compare the predictions from the models across the two wind profile input classes. To answer the second question, we quantitatively compare the model predictions to the measured SCADA power data recorded at the same time as the LiDAR wind profiles used to drive the models. To answer the third question we develop a systematic approach to account for the fact that power scales as the cube of the wind speed. To do this, we normalize the power predictions by the kinetic energy of the inflow wind to remove the first-order effect of wind speed magnitude on power production, thereby revealing the effect of shear and other atmospheric determinants of power production. The following provides a discussion of the results of this study aimed at answering each of the three questions above.

\subsubsection{Canonical ABL profile inputs}\label{sec:Results_Models_Canonical}

\begin{figure*}
\centerline{\includegraphics[width=\textwidth]{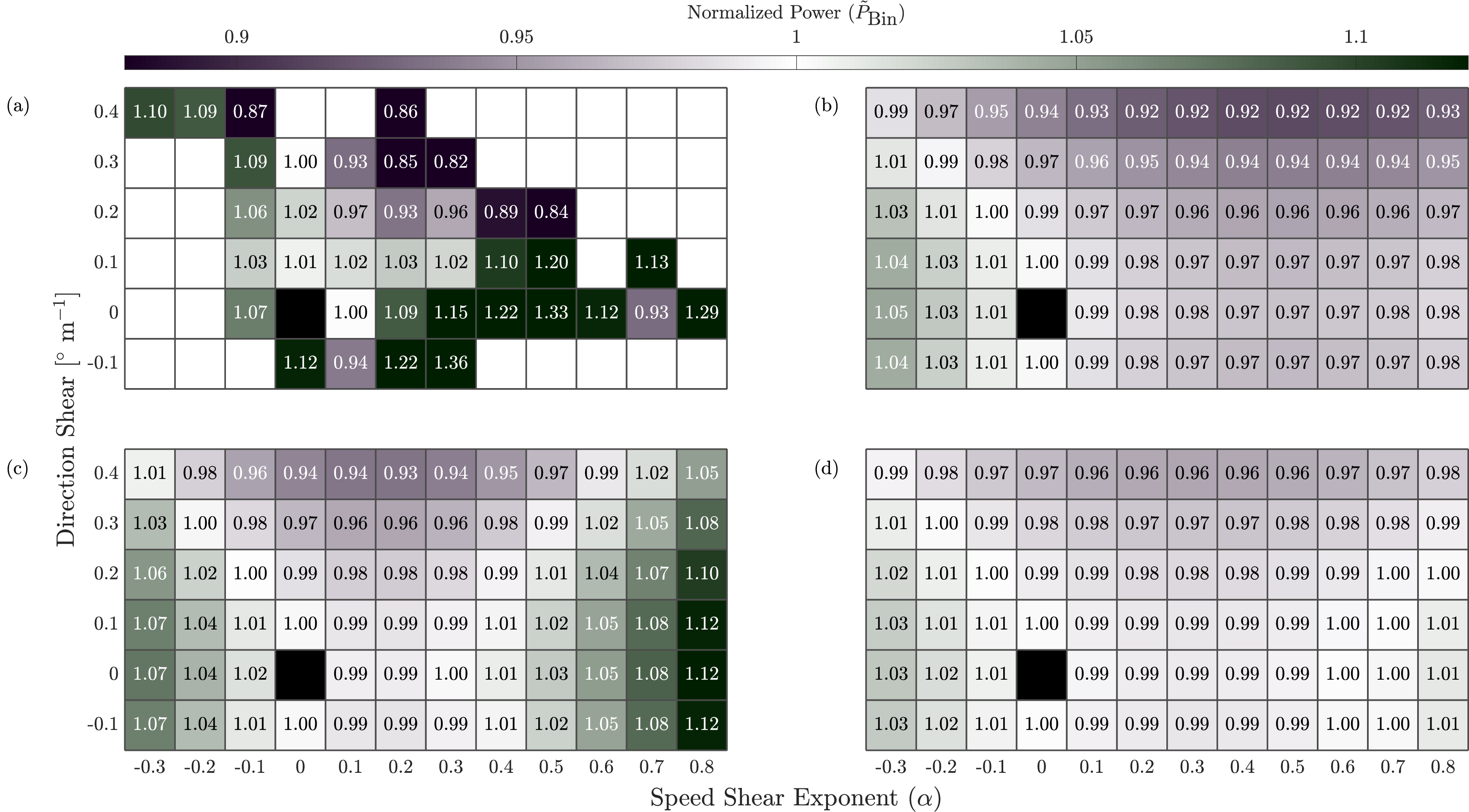}}
\caption{Normalized power ($\tilde{P}_{\text{Bin}}$) with respect to the square representing no speed or direction shear (indicated in black). Panel (a) shows the SCADA power measurements, (b) the REWS model, (c) the REP model, and (d) the BE model from canonical ABL wind profile inputs in which wind speed is modeled with the power law and direction shear is linear over the turbine rotor area. The color axis limits are chosen to highlight trends in the model predictions presented in (b)-(d).\label{fig:Canonical_Contour}}
\end{figure*}

Figs.~\ref{fig:Canonical_Contour} shows the normalized power predictions ($\tilde{P}_{\text{Bin}}$) for the empirical data and each of the three models driven by canonical power law and linear direction shear ABL profile inputs. Each of the three models produces different qualitative and quantitative trends in predicted power over the range of speed and direction shear values considered. The REWS model (Fig.~\ref{fig:Canonical_Contour}(b)) predicts highest performance at the negative speed shear extreme while the REP model (Fig.~\ref{fig:Canonical_Contour}(c)) predicts highest performance at the positive speed shear extreme. The BE model (Fig.~\ref{fig:Canonical_Contour}(d)), in contrast to both REWS and REP, predicts similar levels of over-performance at both positive and negative speed shear extremes, with trends appearing to be approximately symmetric about a speed shear value of approximately $\alpha = 0.3$. We note that the BE model is driven by a constant angular velocity $\Omega$ when driven by the canonical ABL profile inputs.

Each of the three models also shows distinctly different ranges in the spread of predicted over and under-performance. The interval of predicted normalized power for each of the three models is as follows: REWS ($0.92 \leq \tilde{P}_{\text{Bin}} \leq 1.05$), REP ($0.93 \leq \tilde{P}_{\text{Bin}} \leq 1.12$), and BE ($0.96 \leq \tilde{P}_{\text{Bin}} \leq 1.03$). The REP model produces the highest predicted normalized power, 6.5\% higher than the highest performance of the REWS model, and 8.4\% higher than that of the BE model.

It is worth underscoring again that the input wind profiles corresponding to each shear bin in Fig.~\ref{fig:Canonical_Contour}(b)-(d) are identical across each of the three models. Despite this, the power predictions produced by each model show both qualitative and quantitative differences across the range of shear values tested. With respect to the two rotor-equivalent models in Fig.~\ref{fig:Canonical_Contour}(b) and (c), the differences shown highlight the degree of discrepancy induced by computing power from the rotor-equivalent wind speed and the rotor-equivalent wind speed cubed. By placing the cube of the wind speed inside the area integral in Eq.~\ref{eq:P_REP}, an additional region of predicted over-performance is created at the positive speed shear extreme ($\alpha = 0.8$) that the REWS model does not reflect. The BE model shows the least sensitivity to the combinations of speed and direction shear that produce the greatest predicted over and under-performance in the REWS and REP models. While there are two observed areas of predicted over-performance at the extremes of the speed shear axis, similar to the REP model results, the maxima are approximately equal, and much lower in magnitude than those produced by the REP model.

In summary, the two primary results pertaining to the model predictions when driven with canonical ABL profile inputs are: 1) Among the three models, despite being driven with the same profiles, the predicted power across different combinations of speed and direction shear are both qualitatively and quantitatively dissimilar, and 2) the trends in power production for each model are qualitatively and quantitatively distinct from the empirical results observed in Fig.~\ref{fig:Canonical_Contour}(a). In both cases, these differences exist despite the stipulation that the degree of shear in the input wind profiles is classified in the same way (i.e., by Eqs.~\ref{eq:Speed_Shear} and \ref{eq:Dir_Shear}).

\subsubsection{LiDAR inputs}\label{sec:Results_Models_LiDAR}

\begin{figure*}
\centerline{\includegraphics[width=\textwidth]{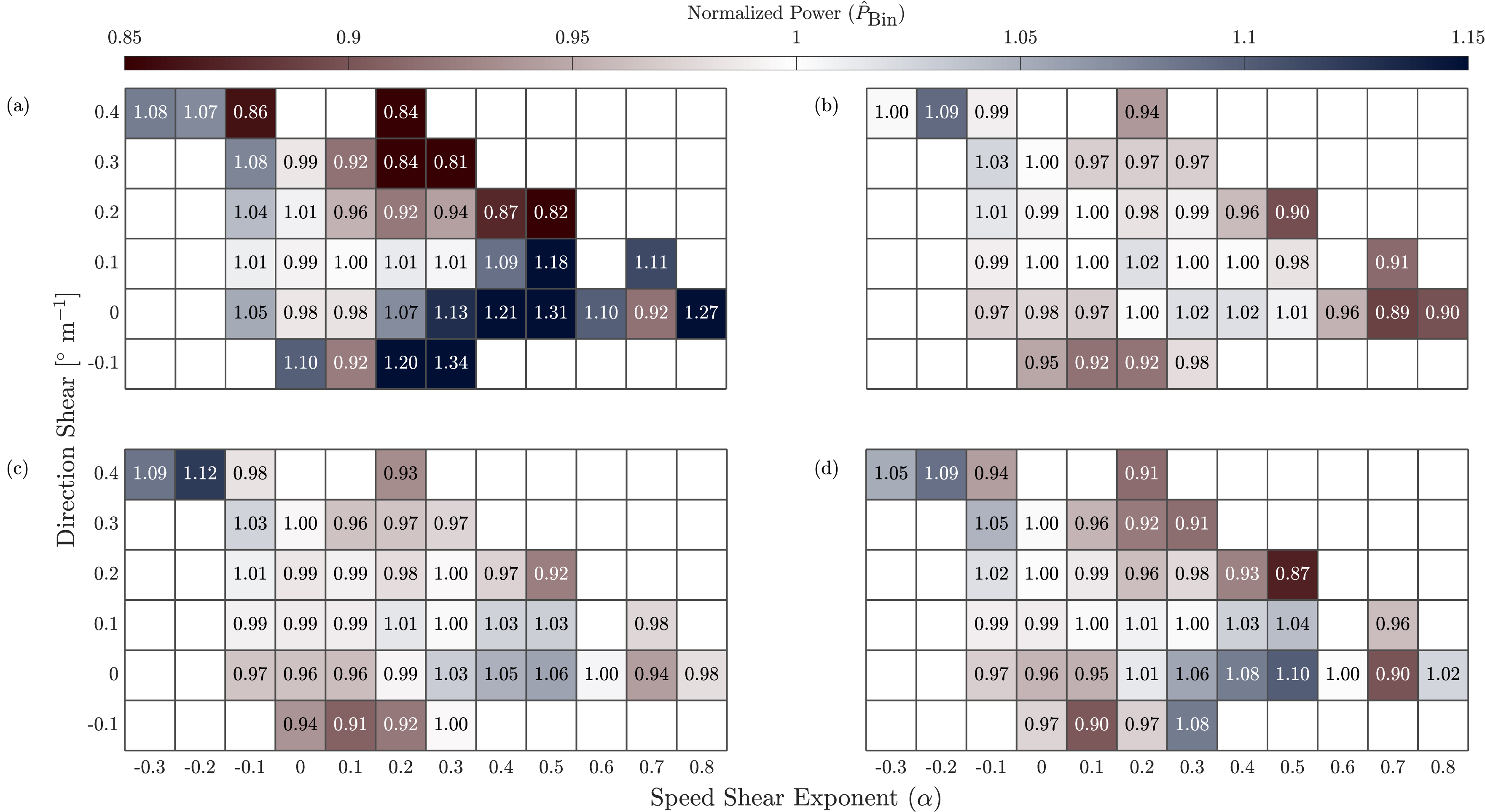}}
\caption{Bin averages of normalized power ($\hat{P}_{\text{Bin}}$) with respect to the median power curve of the turbine at the site for discrete combinations of speed and direction shear. Panel (a) is reproduced from Fig.~\ref{fig:Empirical_Contour}(a) and shows the SCADA power measurements. Panel (b) shows the REWS model, (c) the REP model, and (d) the BE model predictions (Case 1) from LiDAR wind profile inputs. The colormap is chosen to highlight trends in the model predictions presented in (b)-(d).\label{fig:LiDAR_Contour}}
\end{figure*}

\begin{center}
\begin{table*}[t]
\caption{Summary of model correlation and overall error.\label{Correlation_Summary}}
\centering
\begin{tabular*}{500pt}{@{\extracolsep\fill}lcccccc@{\extracolsep\fill}}
\toprule
& \multicolumn{3}{@{}c@{}}{\textbf{$P/P_{\text{Rated}}$}} & \multicolumn{3}{@{}c@{}}{\textbf{$P/P_{\text{Aero}}$}} \\\cmidrule{2-4}\cmidrule{5-7}
\textbf{Model} & $\mathcal{R}$ & \textbf{RMSE} & $\varepsilon_{\text{RMSE}}$ [\%] & $\mathcal{R}$ & \textbf{RMSE} & $\varepsilon_{\text{RMSE}}$ [\%] \\
\midrule
Hub height wind speed       & 0.62 & 0.253 & \textendash & 0.00 & 0.126 & \textendash \\
Rotor-equivalent wind speed & 0.66 & 0.245 & -3.16       & 0.30 & 0.123 & -2.38  \\
Rotor-equivalent power      & 0.68 & 0.239 & -5.53       & 0.34 & 0.120 & -4.76  \\
Blade element (Case 1)      & 0.86 & 0.191 & -24.51      & 0.84 & 0.093 & -26.19 \\
\bottomrule
\end{tabular*}
\begin{tablenotes}
\item A summary of the model correlations and errors relative to measured SCADA power production. The column labeled $P/P_{\text{Rated}}$ shows the model performance when the cubic dependence of power on wind speed dominates other atmospheric determinants of power. The column labeled $P/P_{\text{Aero}}$ shows the model performance when the cubic dependence of power on wind speed is removed by normalizing by the aerodynamic power of the rotor using the hub height wind speed, $P_{\text{Aero}} = \frac{1}{2} \rho A_d U(z_h)^3$. For both cases, the Pearson correlation coefficient $\mathcal{R}$, the root-mean square error (RMSE), and the percent change in RMSE relative to the hub height wind speed model $\varepsilon_{\text{RMSE}}$, are shown.
\end{tablenotes}
\end{table*}
\end{center}

\begin{figure*}
\centerline{\includegraphics[width=4.05in]{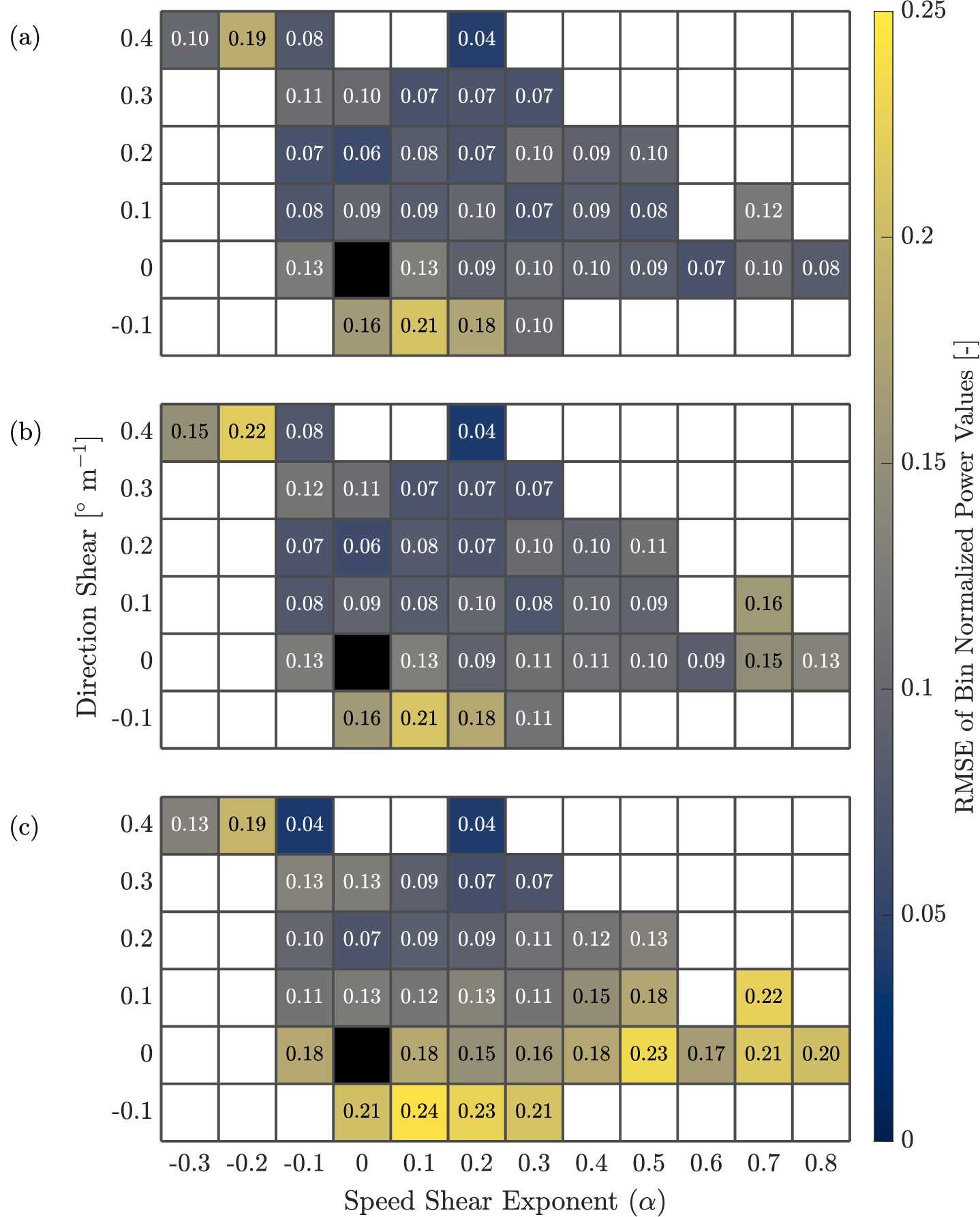}}
\caption{The root mean square error (RMSE) between each of the model outputs from canonical ABL wind profile inputs (see Figs.~\ref{fig:Canonical_Contour}(b)-(d)) and each of the model outputs from LiDAR wind profile inputs (see Figs.~\ref{fig:LiDAR_Contour}(b)-(d)). Panel (a) shows the RMSE for the REWS model, (b) the REP model, and (c) the BE model (Case 1). \label{fig:Canonical_LiDAR_RMSE_Contour}}
\end{figure*}

Fig.~\ref{fig:LiDAR_Contour} shows the empirical results and the model predictions for the LiDAR wind speed and direction measurement inputs. We note that the BE model results here are for Case 1 (constant induction closure of $a=1/3$ and SCADA-measured angular velocity $\Omega$). The BE model results for the other induction closure and controller models are shown in Sec.~\ref{sec:Induction_TSR_Analysis}. As in the empirical results, only bins containing 30 or more points are shown. The trends in power are qualitatively and quantitatively distinct from those produced by the models with canonical ABL profile inputs. The smooth saddle-shape trends observed in the model outputs from canonical ABL wind profile inputs are no longer evident and the spread in predicted over and under-performance has shifted for each model. The REWS model (see Fig.~\ref{fig:LiDAR_Contour}(b)) does not show the characteristic over-performance in the region bounded by $\alpha > 0.1$ and $\beta < \ang{0.2}\text{m}^{-1}$ as expected from the empirical results, which is present in the REP and BE models (Figs.~\ref{fig:LiDAR_Contour}(c) and (d)). The 95\% confidence interval was computed around the mean for each bin, as was done for the empirical results (see Fig.~\ref{fig:Empirical_Contour}(b)). The model predictions, where bins for which the confidence interval includes $\hat{P}_{\text{Bin}} = 1$ have been removed, are shown in the Appendix (see Fig.~\ref{fig:LiDAR_Contour_CI}). The interval of predicted normalized power for each of the three models with LiDAR inputs is as follows: REWS ($0.89 \leq \hat{P}_{\text{Bin}} \leq 1.09$), REP ($0.91 \leq \hat{P}_{\text{Bin}} \leq 1.12$), and BE ($0.87 \leq \hat{P}_{\text{Bin}} \leq 1.10$). One of the primary findings from the results shown in Figs.~\ref{fig:LiDAR_Contour}(b)-(d) is the discrepancy with the power predictions from canonical ABL wind profile inputs in Fig.~\ref{fig:Canonical_Contour}(b)-(d). Even though the values of $\alpha$ and $\beta$ are the same for a given wind profile between the two input classes, the trends in predicted normalized power for each of those two inputs are inconsistent with each other across the range of shear values considered. Fig.~\ref{fig:Canonical_LiDAR_RMSE_Contour} shows the RMSE, defined as

\begin{equation}\label{eq:RMSE_Inputs}
\text{RMSE}_{\text{Inputs}} = \sqrt{\cfrac{1}{N}\sum_{i=1}^N\left[\frac{\hat{P}_i}{\hat{P}_{\text{Bin}}(\beta = \ang{0}\text{m}^{-1},\alpha=0)} - \frac{P^M_{\text{Bin}}}{P^M_{\text{Bin}}(\beta = \ang{0}\text{m}^{-1},\alpha=0)}\right]^2},
\end{equation}

\noindent between the values in each bin for each of the three models with canonical ABL profile inputs. $N$ is the total number of power predictions in each bin from the LiDAR inputs. Differences in the model predictions between the two input wind profile classes tend to be lowest around the region of moderate shear ($\beta = \ang{0.2}\text{m}^{-1}, \alpha = 0.2$). The highest discrepancy occurs in the bin corresponding to $\beta = \ang{-0.1}\text{m}^{-1}$ and $\alpha = 0.1$ for all three models, however the blade element model has two additional regions of high discrepancy located in the area of $\beta = \ang{0}\text{m}^{-1}$ and $\alpha = 0.5$ and $\alpha = 0.7$. Overall, the BE model shows the highest degree of deviation in model predictions between the two input wind profile classes, with a maximum bin RMSE of 0.24, with the REP model next at 0.22, and finally the REWS model at 0.21. It is notable that the largest differences between the canonical and LiDAR input profiles emerge for the BE model. Further, the BE model has the smallest spread in power for canonical inputs, and the largest spread in power for the LiDAR inputs. This suggests that the BE model is more sensitive to nonlinear, non-monotonic variations in the ABL wind profiles, and also suggests the primary impact of the angular velocity $\Omega$. Additional discussion of the variation of rotor angular velocity across different combinations of speed and direction shear, and of the effect of $\Omega$ closure on the BE model results, is provided in Sec.~\ref{sec:Induction_TSR_Analysis}.

\begin{figure*}
\centerline{\includegraphics[width=4in]{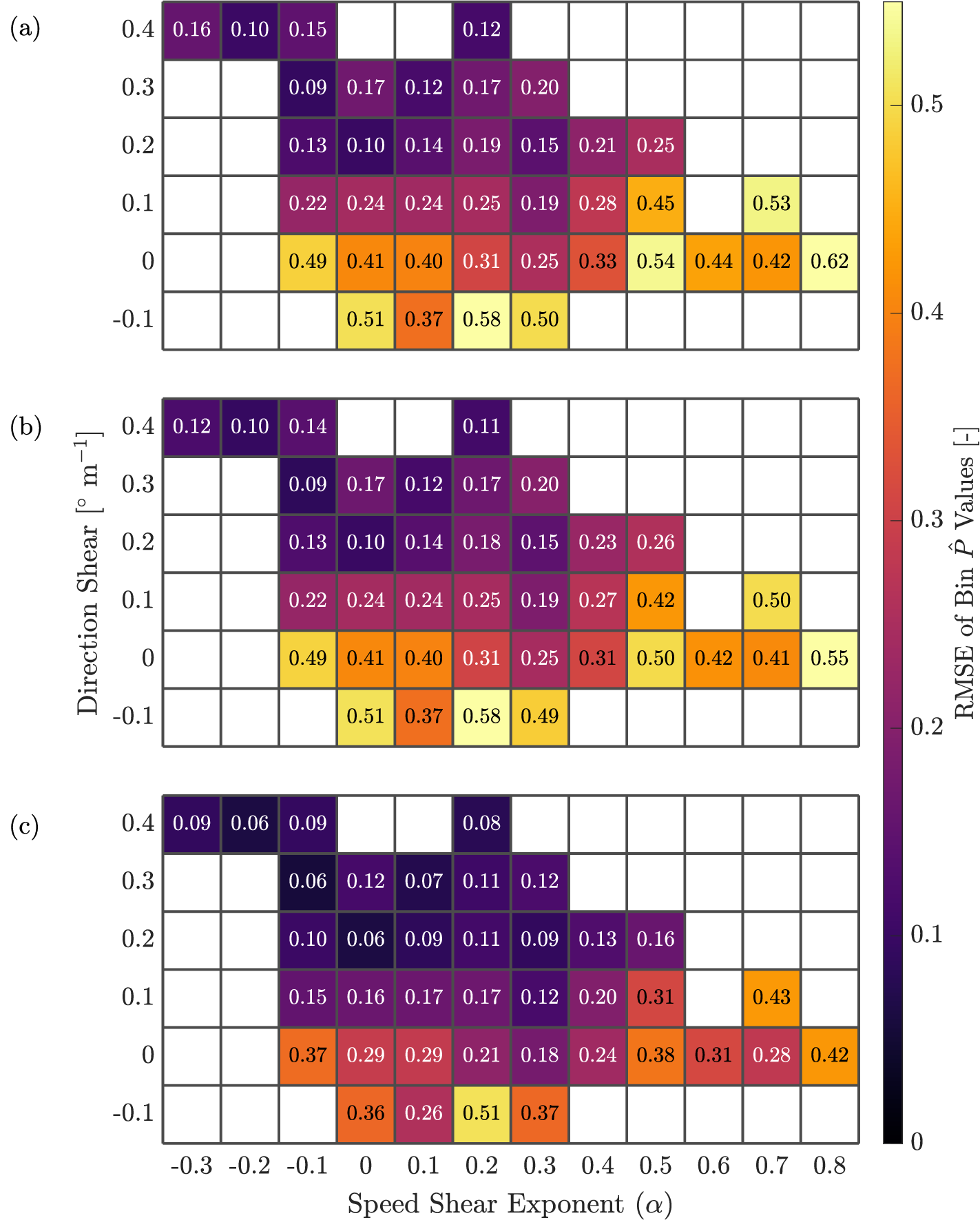}}
\caption{The root mean square error (RMSE) between the empirical values of normalized power (see Fig.~\ref{fig:LiDAR_Contour}(a)) and each of the model outputs from LiDAR wind profile inputs (see Figs.~\ref{fig:LiDAR_Contour}(b)-(d)). Panel (a) shows the RMSE for the REWS model, (b) the REP model, and (c) the BE model (Case 1).\label{fig:Empirical_LiDAR_RMSE_Contour}}
\end{figure*}

We also compare the model predictions from LiDAR inputs Fig.~\ref{fig:LiDAR_Contour}(b)-(d) to the empirical data shown in Fig.~\ref{fig:LiDAR_Contour}(a). Similar to Eq.~\ref{eq:RMSE_Inputs}, the RMSE here is defined as

\begin{equation}\label{eq:RMSE_Empirical}
\text{RMSE}_{\text{Empirical}} = \sqrt{\cfrac{1}{N} \sum^N_{i=1}\left(\hat{P}_{i,\text{Empirical}} - \hat{P}_{i,\text{LiDAR}}\right)^2},
\end{equation}

\noindent and the results are shown in Fig.~\ref{fig:Empirical_LiDAR_RMSE_Contour}. Among the three models, bin RMSE tends to be highest in two separate regions, in the region bounded by $0.1 < \alpha < 0.4$ and $\beta < \ang{0}\text{m}^{-1}$, and in the region bounded by $0.3 < \alpha$ and $-0.1 < \beta < \ang{0.2}\text{m}^{-1}$. Lowest error is observed in the region bounded by $-0.3 < \alpha < 0.3$ and $0.1 < \beta < \ang{0.4}\text{m}^{-1}$. The REWS model produces the highest bin RMSE at 0.62, with the REP model following at 0.58, and finally the BE model producing the overall lowest maximum bin RMSE at 0.51.

\begin{figure*}
\centerline{\includegraphics[width = \textwidth]{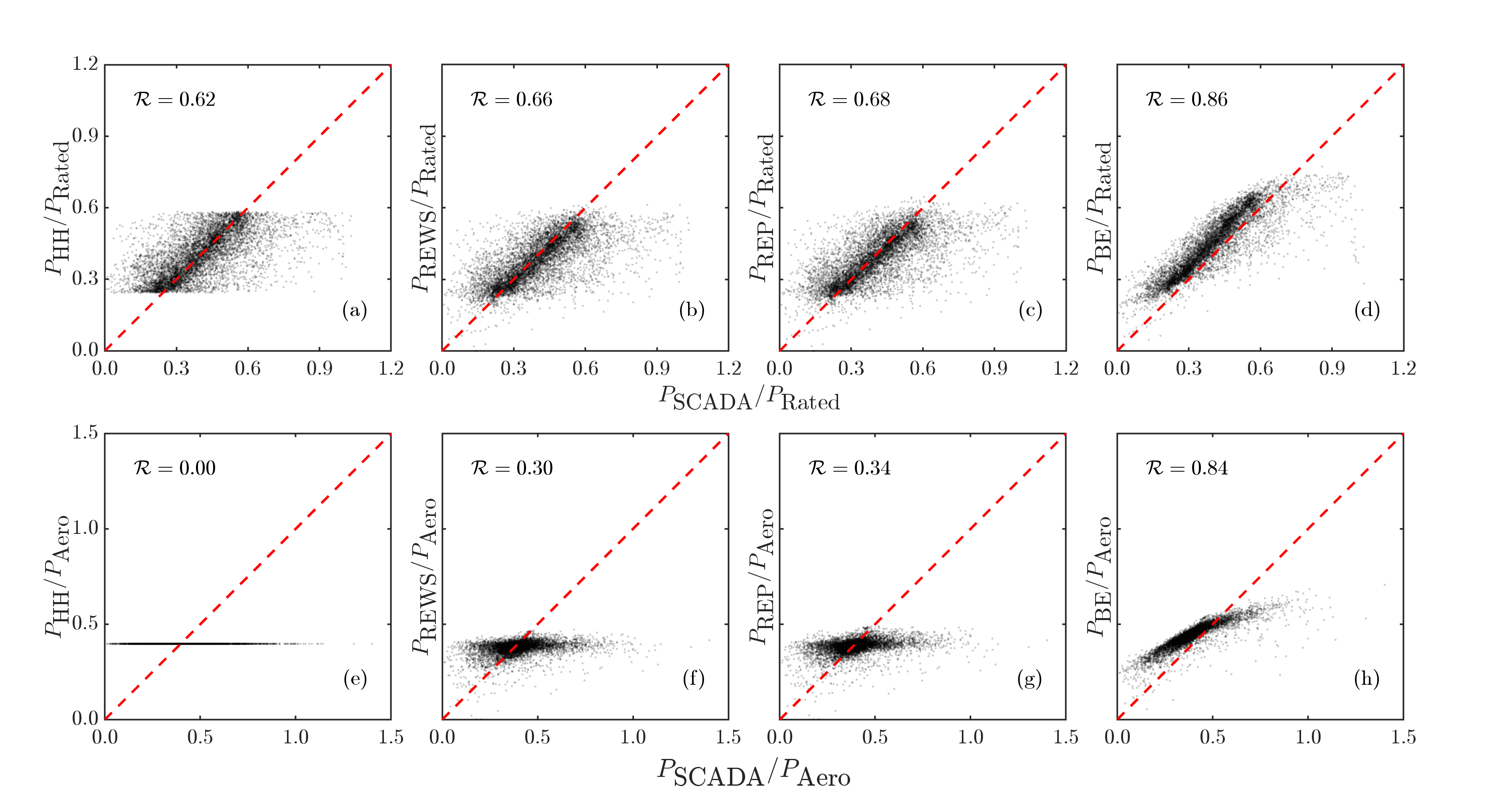}}
\caption{Correlation plots for each of the four models against SCADA power. The top row, (a)-(d), shows power predictions normalized by the rated power of the turbine $P/P_{\text{Rated}}$ and illustrates the performance of the models when the cubic dependence of power on wind speed dominates other atmospheric determinants of power. The bottom row, (e)-(h), shows power predictions normalized by the aerodynamic power of the rotor where the wind speed is measured at hub height, $P_{\text{Aero}} = \frac{1}{2} \rho A_d U(z_h)^3$, and illustrates the performance of the models when the effect of shear and other atmospheric determinants of power are isolated. The Pearson correlation coefficient $\mathcal{R}$ and the root-mean-square error (RMSE) are shown for each case.\label{fig:Correlations}}
\end{figure*}

We show in Fig.~\ref{fig:Correlations} the correlation between individual, 1 min averaged power production predictions for each model driven with LiDAR wind speed profiles compared to the SCADA power measurements from the utility-scale turbine described in Sec.~\ref{sec:Site_Characterization}. Table \ref{Correlation_Summary} provides a concise summary of the Pearson correlation coefficients $\mathcal{R}$ and overall RMSE for each model, as well as the the performance of the models defined as the change in RMSE, $\varepsilon_{\text{RMSE}}$, relative to the hub height wind speed model, described in more detail below.

Fig.~\ref{fig:Correlations}(a)-(d) shows the correlation between model predictions and SCADA data normalized by the rated power of the turbine, a constant that does not depend on the time-varying hub height wind speed. Consequently, the model power predictions and SCADA power measurements are dominated by the cubic dependence of power on wind speed. We discuss first the performance of each model when the effect of wind speed dominates other atmospheric determinants of power, and in the next paragraph, discuss the results when the effect of shear on power production is isolated. Both the rotor-equivalent wind speed and power models show similar correlations ($\mathcal{R}=0.66$ and $0.68$, respectively) and RMSE ($0.245$ and $0.239$, respectively) with SCADA data, with the rotor-equivalent power model having slightly higher correlation and lower error. The blade element model shows the highest correlation ($\mathcal{R}=0.86$) and lowest RMSE ($0.191$) among the four models. The hub height wind speed model, which has no knowledge of shear and is included as a control, shows the lowest correlation ($\mathcal{R}=0.62$) and highest RMSE ($0.253$). Similarly, the blade element model shows the largest reduction in RMSE relative to the hub height wind speed model, $\varepsilon = -24.51\%$, in contrast to $-5.53\%$ for the REP model and $-3.16\%$ for the REWS model. While there is considerable spread between the correlations of the two rotor-equivalent model results and the blade element model results, all three have lower error than the baseline hub height power model.

Fig.~\ref{fig:Correlations}(e)-(h) shows the model predictions and SCADA power correlations non-dimensionalized by the aerodynamic power of the rotor computed using the hub height wind speed from each LiDAR wind speed profile, $P_{\text{Aero}} = \frac{1}{2} \rho A_d U(z_h)^3$. This controls for the cubic dependence of power on wind speed, revealing the effect of shear, among other atmospheric determinants, on power production. Notably, the correlation between the models and the SCADA data significantly decreases when the models do not have the benefit of the cubic dependence of power on wind speed. The REWS and REP models show similar correlations ($\mathcal{R}=0.30$ and $0.34$, respectively) and RMSE ($0.123$ and $0.120$, respectively), with the REP model again showing slightly higher correlation and lower error. Unlike the two rotor-equivalent models, the blade element model correlation remains relatively high ($\mathcal{R}=0.84$) when the cubic dependence of power on wind speed is normalized out, and the RMSE remains the lowest among the models ($0.093$). Again, the blade element model shows a reduction in RMSE relative to the hub height wind speed model ($\varepsilon_{\text{RMSE}} = -26.19\%$) that is 5 to 11 times greater than those of the two rotor-equivalent models ($-2.93\%$ and $-5.22\%$). As confirmation of the theory used to justify the normalization scheme in this analysis, the predictions of the hub height wind speed model in Fig.~\ref{fig:Correlations}(e) are noted to have collapsed to a flat line with no correlation to the SCADA measurements. Because this model has no knowledge of variations in wind speed and direction, when normalized by $P_{\text{Aero}}$, the model predictions should form a line along the value of $C_P$ used in the model. This indicates that the results given for the other models in Fig.~\ref{fig:Correlations}(f)-(h) show the effect of shear on power separated from the cubic dependence of power on wind speed. The results demonstrate that while the REWS and REP models show qualitative differences in their predictions of power based on the inflow shear (the REP model has improved qualitative representation of the dependence of power on wind shear compared to REWS as shown in Fig.~\ref{fig:LiDAR_Contour}), it is challenging to elucidate which actuator disk model representation is more quantitatively accurate in the present study. Further, these results show that with respect to the overarching goal of this study to specifically model the effect of wind speed and direction shear on turbine power production, the blade element model demonstrates a quantitative advantage over the two rotor-equivalent models. 

\subsubsection{Influence of induction closure and controller modeling on blade element model results}\label{sec:Induction_TSR_Analysis}

\begin{figure*}
\centerline{\includegraphics[width = \textwidth]{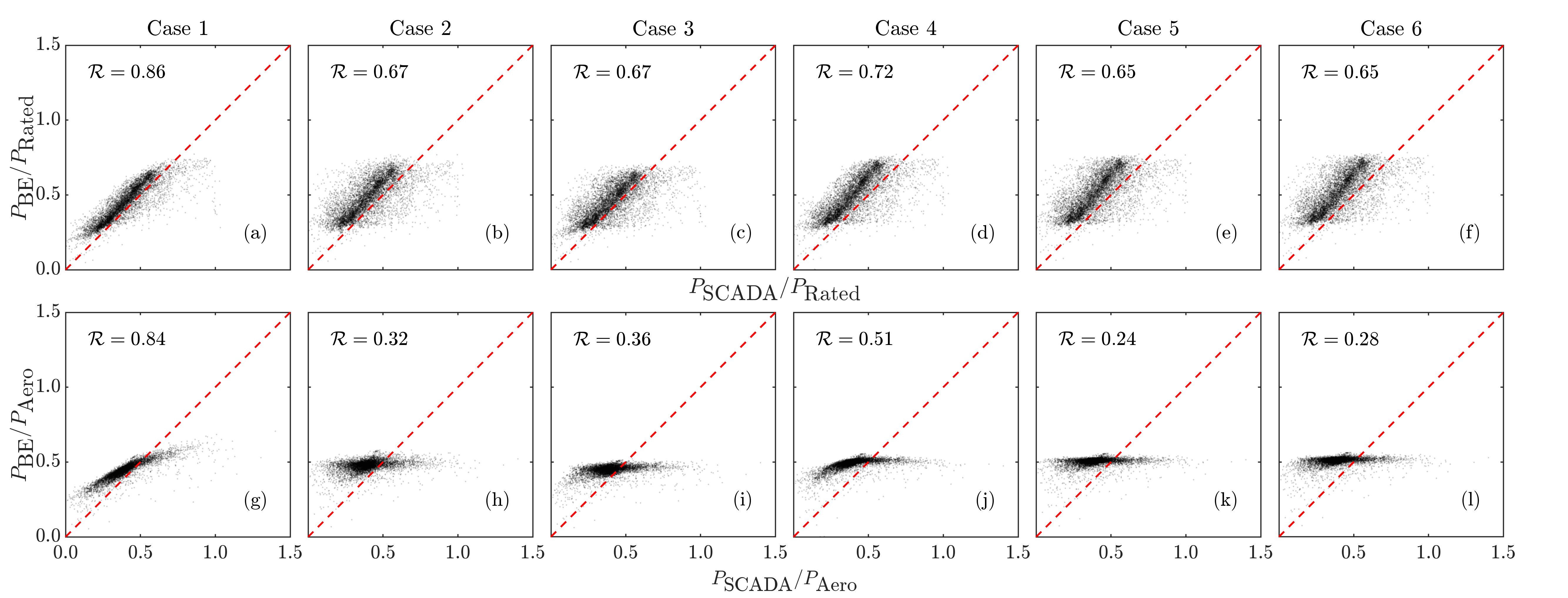}}
\caption{Correlation plots for each of the four sensitivity analysis cases for the BE model with localized induction closure against SCADA power. The top row, (a)-(f), shows power predictions normalized by the rated power of the turbine $P/P_{\text{Rated}}$ and illustrates the performance of the models when the cubic dependence of power on wind speed dominates other atmospheric determinants of power. The bottom row, (g)-(l), shows power predictions normalized by the aerodynamic power of the rotor where the wind speed is measured at hub height, $P_{\text{Aero}} = \frac{1}{2} \rho A_d U(z_h)^3$, and illustrates the performance of the models when the effect of shear and other atmospheric determinants of power are isolated. The Pearson correlation coefficient $\mathcal{R}$ and the root-mean-square error (RMSE) are shown for each case.\label{fig:Sensitivity_Correlations}}
\end{figure*}

In this section we provide the results of the sensitivity analysis on the BE model discussed in Sec.~\ref{sec:Induction_Controller}. Fig.~\ref{fig:Sensitivity_Correlations} shows the correlation between the BE model predictions and SCADA data in the same fashion as Fig.~\ref{fig:Correlations}, for each of the six cases in the sensitivity analysis (two choices for induction closure and three choices for the angular velocity closure yielding six total models). The reference case (Case 1) discussed in Sec.~\ref{sec:Results_Models_LiDAR} where induction is constant and the tip-speed ratio is taken from SCADA data shows the lowest overall error and highest correlation with the empirical data. Among the 5 remaining cases, Case 3 ($a = 1/3; \lambda^*$) demonstrates the lowest overall error, with an increase in RMSE relative to the hub height wind speed model of $2.77\%$ when wind speed dominates predictions of power and $1.59\%$ when the effect of shear is isolated. Case 6 ($a$ from momentum theory; $\lambda^*$) shows the overall highest error and lowest correlation, with an increase in RMSE relative to the hub height wind speed model of $34.78\%$ when wind speed dominates predictions of power and $29.37\%$ when the effect of shear is isolated. The full results of the sensitivity analysis are provided in Table~\ref{Table:sensitivity}.

\begin{figure*}
\centerline{\includegraphics[width=\textwidth]{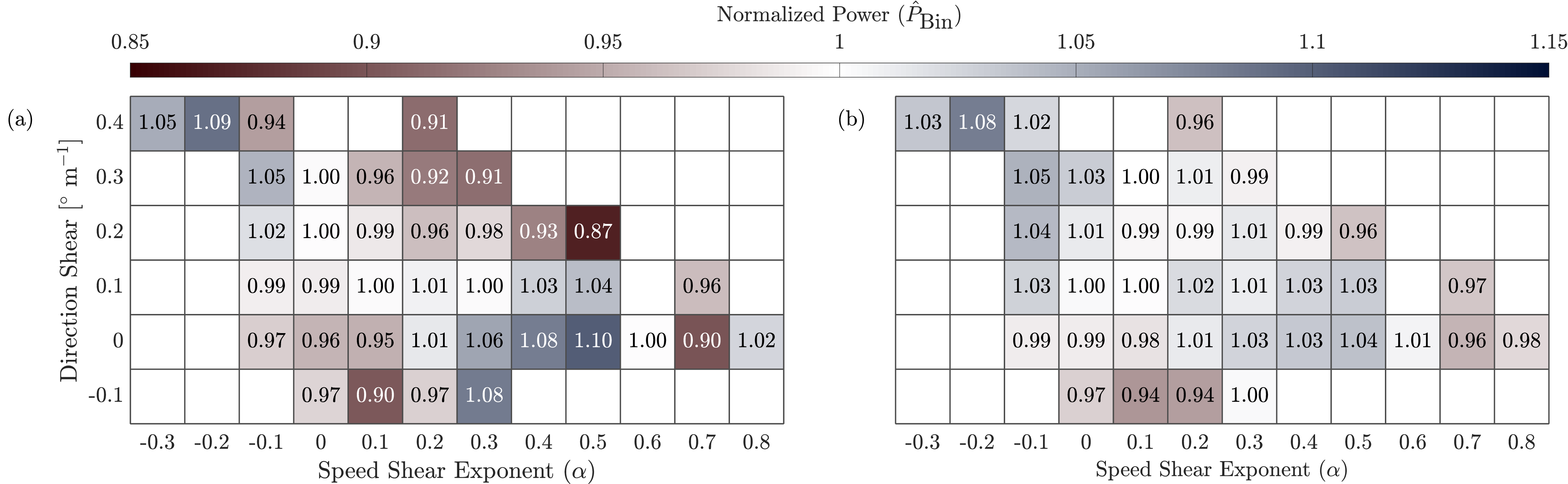}}
\caption{Bin averages of normalized power predictions ($\hat{P}_{\text{Bin}}$) for the BE model. Panel (a) shows the BE model outputs for Case 1 ($a = 1/3; \lambda_{\text{SCADA}}$), the model with the lowest overall error, and (b) the model predictions for Case 3 ($a = 1/3; \lambda^*$), the model setup that produces the lowest overall error when $\Omega$ is not taken from the SCADA data.\label{fig:Effect_of_Omega}}
\end{figure*}

In Fig.~\ref{fig:Effect_of_Omega}, we compare the trends in predicted power production between the best performing setup (Case 1, which uses $\Omega$ from the SCADA data), and the best performing setup in which $\Omega$ is modeled (Case 3). Overall, using the value of $\Omega$ corresponding to $\lambda^*$ rather than $\Omega$ from the SCADA measurements reduces the spread in the maximum degree of predicted over and under-performance. Furthermore, in several bins (e.g., ($\beta = \ang{0.4}\text{m}^{-1}, \alpha = -0.1$) and ($\beta = \ang{0.3}\text{m}^{-1}, \alpha = 0.2$)), predicted under-performance when the BE model is driven with SCADA $\Omega$ becomes predicted over-performance when driven with a constant tip-speed ratio. This comparison, where all other factors are constant between the two cases, demonstrates the degree to which the BE model predictions are dependent on the rotor angular velocity $\Omega$ closure.

\begin{figure*}
\centerline{\includegraphics[width=\textwidth]{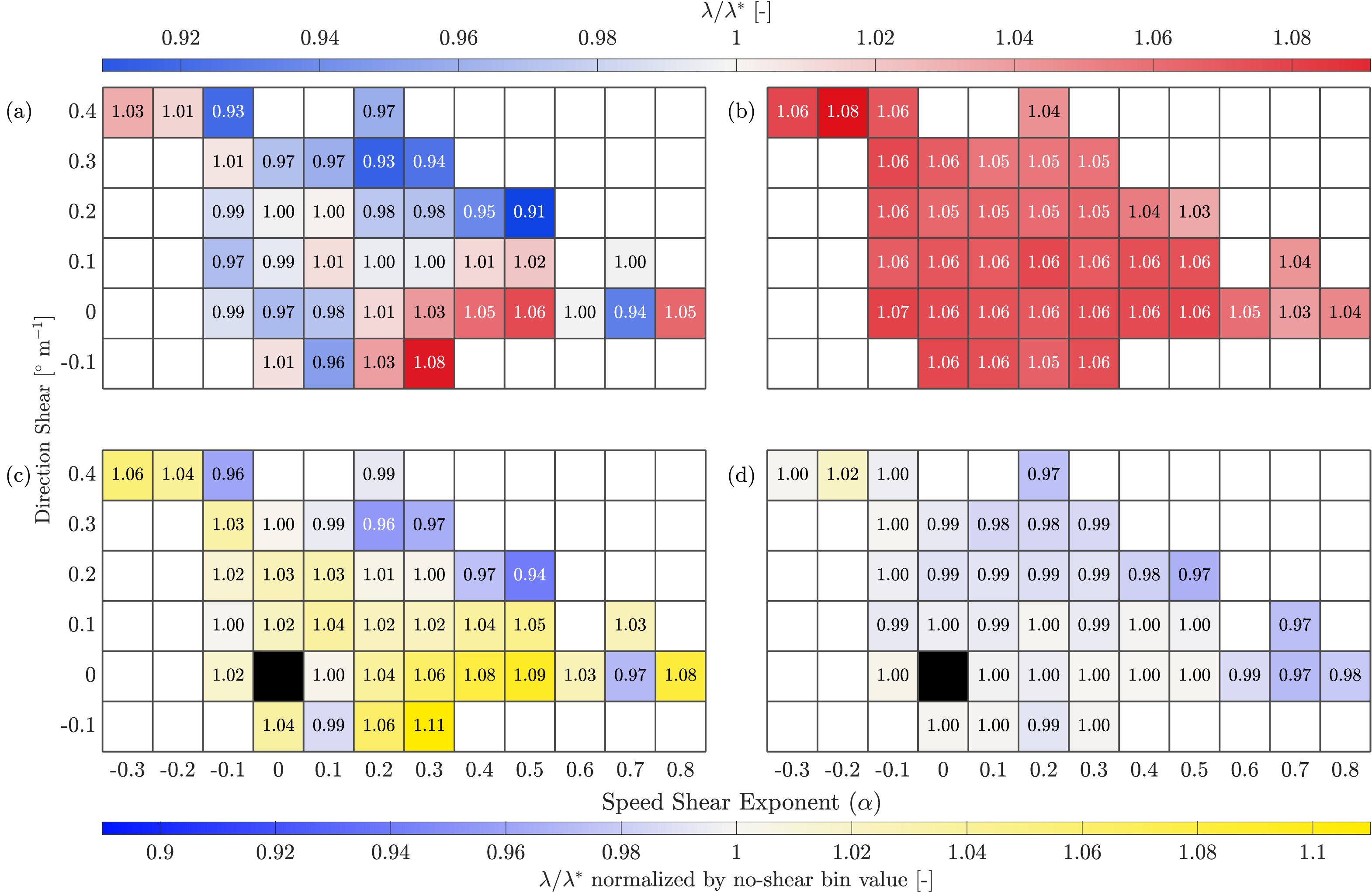}}
\caption{(a) Normalized values of the tip-speed ratio $\lambda$ from the SCADA data. (b) Normalized values of the predicted tip-speed ratio $\lambda$ from the BE model with constant ($a = 1/3$) induction closure and the $k\text{--}\Omega^2$ controller model (Case 2). All values are normalized with respect to the target tip-speed ratio in Region~II ($\lambda^*$). (c) Shows the same information in (a) normalized by the bin value corresponding to no speed or direction shear (indicated by the black square). Similarly, (d) shows the same information in (b) normalized in the same fashion.\label{fig:TSR_Contours}}
\end{figure*}

In Fig.~\ref{fig:TSR_Contours}, we analyze the ability of the $k\text{--}\Omega^2$ model in Case 3 to predict the realized tip-speed ratio. Subplot (a) shows the mean bin value for the realized tip-speed ratio from the SCADA data normalized by the target value commanded by the turbine controller ($\lambda^*$). Subplot (b) shows the mean bin value for the predicted tip-speed ratio from the controller model in Case 3. Subplots (c) and (d) show the same data in (a) and (b), respectively, normalized by the mean value of the bin corresponding to no speed or direction shear. There exist distinct trends in the SCADA data (Fig.~\ref{fig:TSR_Contours}(a)) between above-target tip-speed ratios ($\lambda > \lambda^*$) and over-performance ($\hat{P}_{\text{Bin}} > 1$) in Fig.~\ref{fig:Empirical_Contour}(a)). While there does appear to be some degree of correlation in the model predictions (most noticeable in subplot (d) where normalized $\lambda$ values less than 1 generally coincide with under-performance in Fig.~\ref{fig:Empirical_Contour}(a)), these trends are less pronounced. There also exists less spread in the variation of predicted normalized tip-speed ratios ($0.97$ to $1.02$ in Fig.~\ref{fig:TSR_Contours}(c)) as compared to the SCADA measurements ($0.94$ to $1.11$ in Fig.~\ref{fig:TSR_Contours}(d)). We also observe that, with the exception of a single bin ($\beta = \ang{0.4}\text{m}^{-1}, \alpha = -0.2$), the $k\text{--}\Omega^2$ model universally predicts at or above-target tip-speed ratios ($\lambda \leq \lambda^*$). This demonstrates a potential weakness with implementing the model based on SCADA data. There are inherent differences between the true physical processes taking place at the turbine rotor and the BE model based on the simplifications and assumptions to the model detailed in this section. Imposing a value of $k$ determined from the SCADA data in the BE model operates the turbine away from the true target tip-speed ratio in the field measurements, causing the values of $\lambda/\lambda^*$ in Fig.~\ref{fig:TSR_Contours}(b) to be universally above 1. However, in an analysis where $k$ was chosen to on average produce the target tip-speed ratio from the SCADA data $\lambda^*$, the variations in predicted $\lambda$ (i.e., those in Fig.~\ref{fig:TSR_Contours}(d)) remained unchanged.

\begin{center}
\begin{table*}[t]%
\caption{Summary of correlations and overall error in the BE model sensitivity analysis.\label{BE_Correlation_Summary}}
\centering
\begin{tabular*}{500pt}{@{\extracolsep\fill}lcccccc@{\extracolsep\fill}}
\toprule
& \multicolumn{3}{@{}c@{}}{\textbf{$P/P_{\text{Rated}}$}} & \multicolumn{3}{@{}c@{}}{\textbf{$P/P_{\text{Aero}}$}} \\\cmidrule{2-4}\cmidrule{5-7}
\textbf{BE model setup} & $\mathcal{R}$ & \textbf{RMSE} & $\varepsilon_{\text{RMSE}}$ [\%] & $\mathcal{R}$ & \textbf{RMSE} & $\varepsilon_{\text{RMSE}}$ [\%] \\
\midrule
Case 1 ($a = 1/3; \lambda_{\text{SCADA}}$)      & 0.86 & 0.191 & $-24.51$ & 0.84 & 0.093 & $-26.19$ \\
Case 2 ($a = 1/3; k\text{--}\Omega^2$)          & 0.67 & 0.294 & $+16.21$ & 0.32 & 0.143 & $+13.49$  \\
Case 3 ($a = 1/3; \lambda^*$)                   & 0.67 & 0.260 & $+2.77$  & 0.36 & 0.128 & $+1.59$  \\
Case 4 (mom. theory; $ \lambda_{\text{SCADA}}$) & 0.72 & 0.297 & $+17.39$ & 0.51 & 0.143 & $+13.49$ \\
Case 5 (mom. theory; $k\text{--}\Omega^2$)      & 0.65 & 0.336 & $+32.81$ & 0.24 & 0.160 & $+26.98$ \\
Case 6 (mom. theory; $\lambda^*$)               & 0.65 & 0.341 & $+34.78$ & 0.28 & 0.163 & $+29.37$ \\
\bottomrule
\end{tabular*}
\begin{tablenotes}
\item A summary of the BE model correlations and errors relative to measured SCADA power production. The column labeled $P/P_{\text{Rated}}$ shows the model performance when the cubic dependence of power on wind speed dominates other atmospheric determinants of power. The column labeled $P/P_{\text{Aero}}$ shows the model performance when the cubic dependence of power on wind speed is removed by normalizing by the aerodynamic power of the rotor using the hub height wind speed, $P_{\text{Aero}} = \frac{1}{2} \rho A_d U(z_h)^3$. For both cases, the Pearson correlation coefficient $\mathcal{R}$, the root-mean square error (RMSE), and the percent change in RMSE relative to the hub height wind speed model $\varepsilon_{\text{RMSE}}$, are shown.
\end{tablenotes}
\end{table*}
\end{center}

\section{Discussion and future work}\label{sec:Discussion}

In this study, we analyze a set of LiDAR wind speed and direction measurements and SCADA power measurements to quantify the effect of shear on turbine power production. We observe distinct regions of over and under-performance based on differing combinations of speed and direction shear. We relate these observations to a previous study of similar design~\cite{MSG_2020} and note that different combinations of shear correspond to similar but distinct trends in turbine performance between the two studies. A potential cause of differences between the results of these two studies is the frequency of occurrence of LLJs at each site. As noted in Sec.~\ref{sec:Site_Characterization}, the power law used to characterize the degree of speed shear in both studies is not capable of modeling the complex wind speed profiles inherent to LLJs. Specifically, the power law cannot model wind speed profiles that do not increase or decrease monotonically with height. When the core of a LLJ descends into the turbine rotor area, there exists some degree of both positive and negative shear over the rotor, which the power law cannot characterize. This may increase the error between the power law fit and the true wind speed profile, which could shift data between bins in the plots shown in Figs.~\ref{fig:Canonical_Contour} and \ref{fig:LiDAR_Contour}, thereby changing the observed empirical trends in over and under-performance.

Given these limitations in parsing the empirical trends based on the high-dimensional ABL flow, we focus on developing and assessing parametric models for the effect of wind shear on power production. We assess the predictions of three different models for turbine power production that account for wind speed and direction shear over the rotor area, and compare these results to a model that has no knowledge of wind shear. We drive each model with two separate and distinct input classes: 1) best-fit canonical ABL profiles where wind speed is modeled with the power law and wind direction varies linearly with height, and 2) wind speed and direction measurements made by a profiling LiDAR. These two input types are observed to produce qualitative and quantitative differences in power predictions in each model. These results indicate that the variations present in complex wind speed and direction profiles that are not present in median profiles have a first-order effect on model outputs. This implies that the model outputs may vary substantially as the frequency of the input wind conditions changes. That is, as the time averaging window of the input wind profiles increases, for instance to 10-min, as is standard in industry wind resources assessments, the model predictions may vary from those produced by higher frequency inputs like those used in this study. Since power production is a nonlinear function of the wind speed, the average power production based on instantaneous wind is not equivalent to the power production based on the averaged wind, and there is therefore inherently a dependence on the time averaging scale. Future work should further elucidate the time averaging scale that is necessary to accurately model the influence of wind shear, since lowering the averaging period will also further complicate the qualitative characterization of the profiles as they become more sensitive to microscale ABL turbulence. Further, the present results have an implication on wind resource assessment: power predictions assuming power law profile behavior made in the absence of real wind profile measurements may deviate substantially from true values based on the assumptions made about the input wind profiles. Specifically, if canonical ABL profiles are used in which no short-term turbulent fluctuations, or diurnally-varying atmospheric features (such as LLJs) are accounted for, predictions of power may not accurately reflect the true power production at a given site.

We compare the predictions from each model driven with LiDAR wind speed and direction inputs to concurrent power measurements from a utility-scale wind turbine located adjacent to the LiDAR system. The hub height wind speed model has no knowledge of variations in wind speed and direction over the rotor and is included as a reference against which the other models that do have knowledge of variations over the rotor are compared. We assess the models in two separate ways. First, the model results are analyzed when power predictions are dominated by the wind speed magnitude in the input wind profile. Second, we isolate the effect of shear and other determinants of power by normalizing the power predictions by the aerodynamic power of the rotor based on the hub height wind speed of the input wind profile. In both cases, we observe that the blade element model shows the highest correlation and lowest overall error with the SCADA measurements. The two rotor-equivalent models show similar correlations and overall errors in both cases. However, the correlations of the REWS and REP models to the SCADA measurements decrease by a factor of 2 when normalized to isolate the effect of shear, whereas the blade element model correlation decreases only marginally. The implication of this is that the blade element model is more able to accurately model the effect of speed and direction shear in arbitrary wind profile inputs on turbine power production than the rotor-equivalent models in their current form. 

While the results presented in Fig.~\ref{fig:Correlations} show promise to capture the effects of wind shear with a blade element approach, a notable weakness for this model is the additional information required, which is not necessary to drive the two rotor-equivalent models. Namely, the airfoil properties corresponding to each blade node, as well as the rotor angular velocity corresponding to the atmospheric conditions in each input wind profile. Fig.~\ref{fig:Effect_of_Omega} demonstrates the degree to which knowledge of the angular velocity affects the BE model power predictions. When $\Omega$ is computed from the target tip-speed ratio of the turbine $\lambda^*$, the BE model accuracy decreases substantially. The correlation with SCADA power measurements when power is dominated by the cubic dependence of power on wind speed  falls to $\mathcal{R} = 0.67$, and when the cubic dependence is accounted for, the correlation is $\mathcal{R} = 0.36$. With knowledge of $\Omega$ removed, in the best case, results are comparable to the two rotor-equivalent models, and in the remaining cases under-perform even the hub height wind speed model. Moreover, Fig.~\ref{fig:TSR_Contours}(a) shows that the average tip-speed ratio for discrete combinations of speed and direction shear follows similar trends as the normalized power production shown in Fig.~\ref{fig:LiDAR_Contour}. This suggests that the BE model depends on information pertaining to the specific operating strategy commanded by the onboard control system, and when this information is removed by assuming a constant tip-speed ratio for Region~II operation, rather than known a priori, the BE model has similar performance to the actuator disc REWS and REP models considered. Therefore, future work on accurately estimating the tip-speed ratio depending on wind shear, and other atmospheric and wind turbine control determinants, is particularly pertinent.

Uncertainty exists around how rotor aerodynamics (e.g., induction factors) are affected by wind speed and direction shear. We test several simplifications and assumptions to demonstrate their effect on the BE model results. Elucidating how these aerodynamic mechanisms depend on wind speed and direction shear, and then incorporating any corresponding extensions developed into the BE framework proposed here, is suggested for future work. In a sensitivity test of the effect of induction closure and turbine control on the BE model results, we find that the model is dependent on knowledge of the rotor speed to first order. Further, when the tip-speed ratio is estimated using the standard controller model from the literature ($k\text{--}\Omega^2$), the predictions are under-dispersive relative to the measured SCADA values. This result holds regardless of the induction closure used, however, induction closure from one-dimensional momentum theory appears to further increase error relative to the measured SCADA data. This indicates that the momentum theory induction closure model has a stabilizing effect on the controller model, such that when both are used in tandem, feedback between the two models tends to mitigate the effect of arbitrary wind speed and direction shear on model outputs. Moreover, none of the cases tested in the sensitivity analysis were able to meet or exceed the benchmark correlation or error achieved by the BE model when run with constant induction closure and the SCADA tip-speed ratio. This again underscores the impact that knowledge of the tip-speed ratio has on the BE model predictions and denotes limitations in both the induction closure and controller models in handling inputs containing shear. Future work should focus on understanding how the induced velocities depend on wind speed and direction shear, and on estimating $\lambda$, and by extension $\Omega$, based on the turbine controller response to inflow wind conditions. This will enable the BE model to be used for predictions where SCADA measurements of these quantities are not available (e.g., wind farm siting and design).

Finally, the BE model does not consider the effect of any other determinants of turbine efficiency, such as turbulence at scales smaller than the time averaging timescale, which has been studied in empirical investigations and actuator disc formulations extensively (see, e.g., Wharton and Lundquist~\cite{Wharton_2012a,Wharton_2012b}, Murphy et al.~\cite{Murphy_2020}, Brand et al.~\cite{Brand_2011}, Ryu et al.~\cite{Ryu_2022}, St. Martin et al.~\cite{StMartin_2016}). While high frequency turbulence measurements over the rotor area were not available in this study, future work may assess the potential joint role of shear and turbulence using meteorological tower measurements and computational fluid dynamics. In this study, the blade element model has lowest error in predicting the power depending on wind shear both with and without knowledge from SCADA measurements. The model that is able to most accurately predict turbine power production depending on wind speed and direction shear, without knowledge from SCADA measurements, across multiple wind farm sites should be incorporated into wind resource assessments for wind farm siting and design. 


\section*{Acknowledgments}

S.A.M. gratefully acknowledges partial support from the MIT Office of the Vice Chancellor, the Gates Millennium Scholars Program, Chevron Corporation, and the United States Department of Energy. M.F.H. gratefully acknowledges support from the National Science Foundation under grant no. CBET 2226053 (Program Manager Ron Joslin) and partial support from Siemens Gamesa Renewable Energy.

This material is based upon work supported by the U.S. Department of Energy, Office of Science, Office of Advanced Scientific Computing Research, Department of Energy Computational Science Graduate Fellowship under Award Number DE-SC0023112.

This report was prepared as an account of work sponsored by an agency of the United States Government. Neither the United States Government nor any agency thereof, nor any of their employees, makes any warranty, express or implied, or assumes any legal liability or responsibility for the accuracy, completeness, or usefulness of any information, apparatus, product, or process disclosed, or represents that its use would not infringe privately owned rights. Reference herein to any specific commercial product, process, or service by trade name, trademark, manufacturer, or otherwise does not necessarily constitute or imply its endorsement, recommendation, or favoring by the United States Government or any agency thereof. The views and opinions of authors expressed herein do not necessarily state or reflect those of the United States Government or any agency thereof.

The authors thank Jaime Liew and Paul Hulsman for many productive discussions throughout the preparation of this report. The authors would also like to thank ReNew Power, Siemens Gamesa Renewable Energy, and Varun Sivaram for support throughout this study.

\subsection*{Author contributions}

Storm A. Mata: Investigation (lead); Visualization (lead); Writing - original draft preparation (lead); Writing - review \& editing (equal). Michael F. Howland: Conceptualization (lead); Supervision (lead); Writing - review \& editing (equal). Juan Jos{\'e} Pena Mart{\'i}nez, Jes{\'u}s Bas Quesada, Felipe Palou Larra\~{n}aga, Neeraj Yadav, Jasvipul S. Chawla, and Varun Sivaram: Data collection and curation (equal).

\subsection*{Financial disclosure}

None reported.

\subsection*{Conflict of interest}

The authors declare no potential conflict of interests.

\appendix

\section{The Statistical Significance of Power Predictions}
\label{app1}

\begin{figure*}
\centerline{\includegraphics[width=\textwidth]{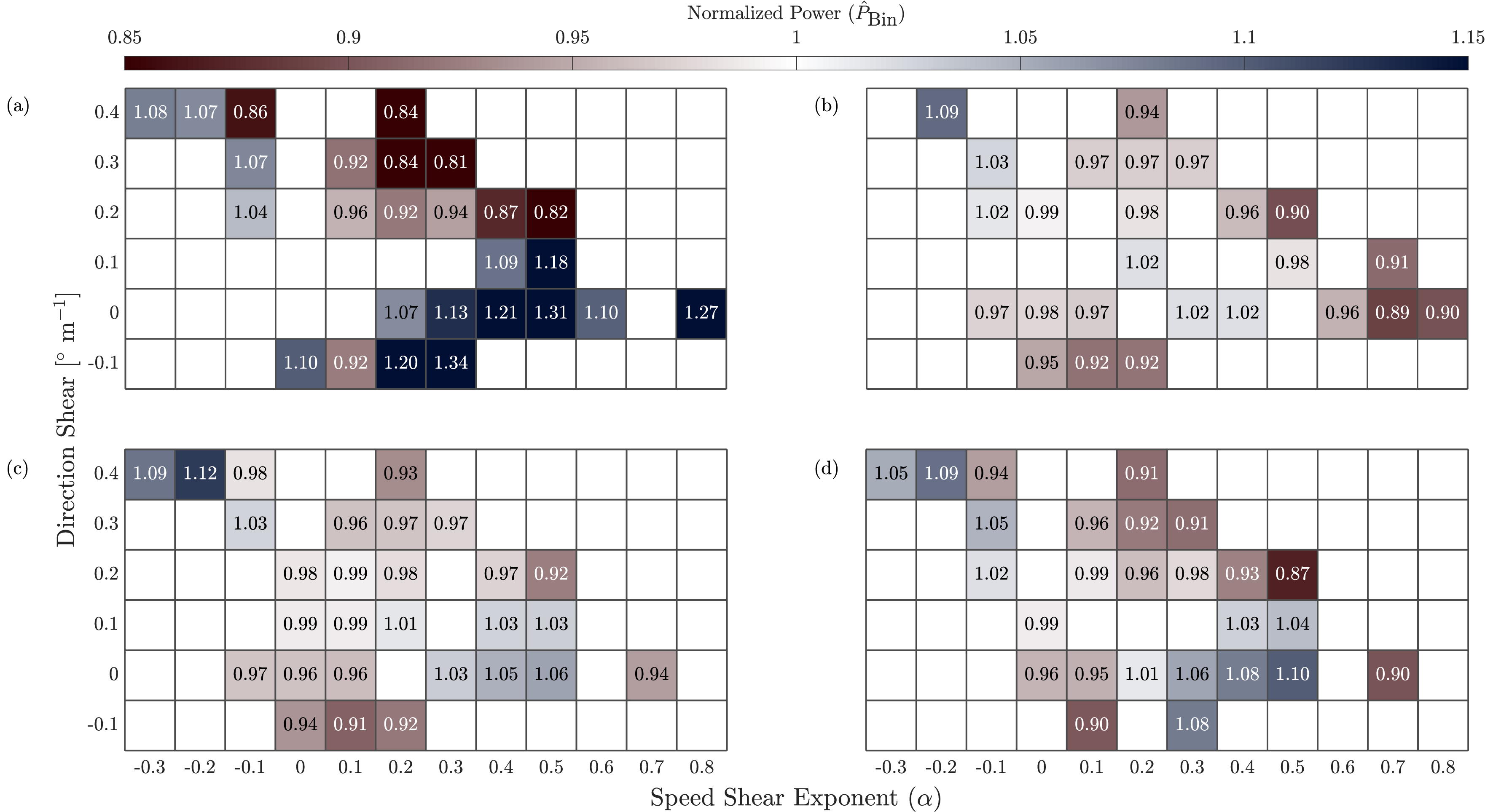}}
\caption{Bin averages of normalized power ($\hat{P}_{\text{Bin}}$) with respect to the median power curve of the turbine at the site for discrete combinations of speed and direction shear. Panel (a) shows the SCADA power measurements. Panel (b) shows the REWS model predictions, (c) the REP model, and (d) the BE model from LiDAR wind profile inputs. The colormap is chosen to highlight trends in the model predictions presented in (b)-(d). The 95\% confidence interval (CI) is computed for each bin. Bins for which the CI spans the value of $\hat{P}_{\text{Bin}} = 1$ are removed because trends in over-and under-performance indicated in these bins cannot be discerned reliably with the given data.\label{fig:LiDAR_Contour_CI}}
\end{figure*}

To determine the statistical significance of the trends in over and under-performance in the empirical results and model predictions, we perform the following analysis. The 95\% confidence interval (CI) is computed around the mean value of each bin shown in Fig.~\ref{fig:LiDAR_Contour} using bootstrapping. In Fig.~\ref{fig:LiDAR_Contour_CI}, we remove the bins in which the CI contains the value of normalized power equal to one ($\hat{P}_{\text{Bin}} = 1$). These bins are removed because the CI crosses the threshold of over and under-performance, indicating that at the selected 95\% significance level, the performance reflected in these bins is not statistically significant.

\FloatBarrier

\bibliography{wileyNJD-AMA}%

\begin{thebibliography}{10}
\providecommand \doibase [0]{http://dx.doi.org/}%

\bibitem{Land_Based_2022}
Wiser R, Bolinger M, Hoen B, et al. Land-Based Wind Market Report: 2022 Edition. Online;  2022.

\bibitem{Offshore_2022}
Musial W, Spitsen P, Duffy P, et al. Offshore Wind Market Report: 2022 Edition. Online;  2022.

\bibitem{Ryu_2022}
Ryu GH, Kim D, Kim DY, et al. Analysis of Vertical Wind Shear Effects on Offshore Wind Energy Prediction Accuracy Applying Rotor Equivalent Wind Speed and the Relationship with Atmospheric Stability. {\it Applied Sciences} 2022\string; 12(14).
\newblock \href {\doibase 10.3390/app12146949} {doi: 10.3390/app12146949}

\bibitem{Veers_2019}
Veers P, Dykes K, Lantz E, et al. Grand challenges in the science of wind energy. {\it Science} 2019\string; 366(6464)\string: eaau2027.

\bibitem{Stull_ch1}
Stull R. {\it Mean Boundary Layer Characteristics}\string: 1--28; Dordrecht, The Netherlands: Kluwer Academic Publishers .
\newblock 1988.

\bibitem{lange2017wind}
Lange J, Mann J, Berg J, et al. For wind turbines in complex terrain, the devil is in the detail. {\it Environmental Research Letters} 2017\string; 12(9)\string: 094020.

\bibitem{liew2020analytical}
Liew J, Urb{\'a}n AM, Andersen SJ. Analytical model for the power--yaw sensitivity of wind turbines operating in full wake. {\it Wind Energy Science} 2020\string; 5(1)\string: 427--437.

\bibitem{Lundquist_2020}
Lundquist JK. Wind Shear and Wind Veer Effects on Wind Turbines. In:  Stoevesandt B, Schepers G, Fuglsang P, Yuping S. \kern-2pt, eds. {\it Handbook of Wind Energy Aerodynamics}Cham, Switzerland: Springer International Publishing.  2020 (pp. 1--22)

\bibitem{Wyngaard_ch9}
Wyngaard JC. {\it The atmospheric boundary layer}\string: 193--2014; Cambridge, United Kingdom: Cambridge University Press .
\newblock 2010.

\bibitem{Debnath_2023}
Debnath M, Moriarty P, Krishnamurthy R, et al. Characterization of wind speed and directional shear at the awaken field campaign site. {\it Journal of Renewable and Sustainable Energy} 2023\string; 15(3).

\bibitem{Walter_2009}
Walter K, Weiss CC, Swift AHP, Chapman J, Kelley ND. {Speed and Direction Shear in the Stable Nocturnal Boundary Layer}. {\it Journal of Solar Energy Engineering} 2009\string; 131(1)\string: 011013.
\newblock \href {\doibase 10.1115/1.3035818} {doi: 10.1115/1.3035818}

\bibitem{howland2022optimal}
Howland MF, Ghate AS, Quesada JB, et al. Optimal closed-loop wake steering--Part 2: Diurnal cycle atmospheric boundary layer conditions. {\it Wind Energy Science} 2022\string; 7(1)\string: 345--365.

\bibitem{Van_Ulden1985}
Ulden APV, Holtslag AAM. Estimation of Atmospheric Boundary Layer Parameters for Diffusion Applications. {\it Journal of Applied Meteorology and Climatology} 1985\string; 24(11)\string: 1196 - 1207.
\newblock \href {\doibase https://doi.org/10.1175/1520-0450(1985)024<1196:EOABLP>2.0.CO;2} {doi: https://doi.org/10.1175/1520-0450(1985)024<1196:EOABLP>2.0.CO;2}

\bibitem{Pena2014}
Alfredo~Peña SEG. The H{\o}vs{\o}re Tall Wind-Profile Experiment: A Description of Wind Profile Observations in the Atmospheric Boundary Layer. {\it Boundary-Layer Meteorology} 2014\string; 150(1)\string: 69-89.
\newblock \href {\doibase https://doi.org/10.1007/s10546-013-9856-4} {doi: https://doi.org/10.1007/s10546-013-9856-4}

\bibitem{Lindvall2019}
Lindvall J, Svensson G. Wind turning in the atmospheric boundary layer over land. {\it Quarterly Journal of the Royal Meteorological Society} 2019\string; 145(724)\string: 3074-3088.
\newblock \href {\doibase https://doi.org/10.1002/qj.3605} {doi: https://doi.org/10.1002/qj.3605}

\bibitem{Shu_2020}
Shu Z, Li Q, He Y, Chan PW. Investigation of Marine Wind Veer Characteristics Using Wind Lidar Measurements. {\it Atmosphere} 2020\string; 11(11).
\newblock \href {\doibase 10.3390/atmos11111178} {doi: 10.3390/atmos11111178}

\bibitem{Wharton_2012a}
Wharton S, Lundquist JK. Assessing atmospheric stability and its impacts on rotor-disk wind characteristics at an onshore wind farm. {\it Wind Energy} 2012\string; 15(4)\string: 525-546.
\newblock \href {\doibase https://doi.org/10.1002/we.483} {doi: https://doi.org/10.1002/we.483}

\bibitem{Wharton_2012b}
Wharton S, Lundquist JK. Atmospheric stability affects wind turbine power collection. {\it Environmental Research Letters} 2012\string; 7(1)\string: 014005.
\newblock \href {\doibase 10.1088/1748-9326/7/1/014005} {doi: 10.1088/1748-9326/7/1/014005}

\bibitem{StMartin_2016}
St.~Martin CM, Lundquist JK, Clifton A, Poulos GS, Schreck SJ. Wind turbine power production and annual energy production depend on atmospheric stability and turbulence. {\it Wind Energy Science} 2016\string; 1(2)\string: 221--236.
\newblock \href {\doibase 10.5194/wes-1-221-2016} {doi: 10.5194/wes-1-221-2016}

\bibitem{Vanderwende_2012}
Vanderwende BJ, Lundquist JK. The modification of wind turbine performance by statistically distinct atmospheric regimes. {\it Environmental Research Letters} 2012\string; 7(3)\string: 034035.
\newblock \href {\doibase 10.1088/1748-9326/7/3/034035} {doi: 10.1088/1748-9326/7/3/034035}

\bibitem{Murphy_2020}
Murphy P, Lundquist JK, Fleming P. How wind speed shear and directional veer affect the power production of a megawatt-scale operational wind turbine. {\it Wind Energy Science} 2020\string; 5(3)\string: 1169--1190.
\newblock \href {\doibase 10.5194/wes-5-1169-2020} {doi: 10.5194/wes-5-1169-2020}

\bibitem{Howland_2020}
Howland MF, {Moral González} C, {Pena Martínez} JJ, et al. {Influence of atmospheric conditions on the power production of utility-scale wind turbines in yaw misalignment}. {\it Journal of Renewable and Sustainable Energy} 2020\string; 12(6).
\newblock 063307.\href {\doibase 10.1063/5.0023746} {doi: 10.1063/5.0023746}

\bibitem{howland2022collective}
Howland MF, Quesada JB, Mart{\'\i}nez JJP, et al. Collective wind farm operation based on a predictive model increases utility-scale energy production. {\it Nature Energy} 2022\string; 7(9)\string: 818--827.

\bibitem{MSG_2020}
Sanchez~Gomez M, Lundquist JK. The effect of wind direction shear on turbine performance in a wind farm in central Iowa. {\it Wind Energy Science} 2020\string; 5(1)\string: 125--139.
\newblock \href {\doibase 10.5194/wes-5-125-2020} {doi: 10.5194/wes-5-125-2020}

\bibitem{Manwell_Ch2}
Manwell JF, McGowan JG. Wind Characteristics and Resources. In: New Jersey, USA: John Wiley \& Sons, Ltd.  2009 (pp. 23--90).

\bibitem{Lydia_2013}
Lydia M, Selvakumar AI, Kumar SS, Kumar GEP. Advanced Algorithms for Wind Turbine Power Curve Modeling. {\it IEEE Transactions on Sustainable Energy} 2013\string; 4(3)\string: 827-835.
\newblock \href {\doibase 10.1109/TSTE.2013.2247641} {doi: 10.1109/TSTE.2013.2247641}

\bibitem{Lydia_2014}
Lydia M, Kumar SS, Selvakumar AI, {Prem Kumar} GE. A comprehensive review on wind turbine power curve modeling techniques. {\it Renewable and Sustainable Energy Reviews} 2014\string; 30\string: 452-460.
\newblock \href {\doibase https://doi.org/10.1016/j.rser.2013.10.030} {doi: https://doi.org/10.1016/j.rser.2013.10.030}

\bibitem{Wagner_2009}
Wagner R, Antoniou I, Pedersen SM, Courtney MS, Jørgensen HE. The influence of the wind speed profile on wind turbine performance measurements. {\it Wind Energy} 2009\string; 12(4)\string: 348-362.
\newblock \href {\doibase https://doi.org/10.1002/we.297} {doi: https://doi.org/10.1002/we.297}

\bibitem{Choukulkar_2016}
Choukulkar A, Pichugina Y, Clack CTM, et al. A new formulation for rotor equivalent wind speed for wind resource assessment and wind power forecasting. {\it Wind Energy} 2016\string; 19(8)\string: 1439-1452.
\newblock \href {\doibase https://doi.org/10.1002/we.1929} {doi: https://doi.org/10.1002/we.1929}

\bibitem{Wagner_2011}
Wagner R, Courtney MS, Gottschall J, Lindel\"ow PJP. Accounting for the speed shear in wind turbine power performance measurement. {\it Wind Energy} 2011\string; 14(8)\string: 993-1004.
\newblock \href {\doibase 10.1002/we.509} {doi: 10.1002/we.509}

\bibitem{Clack_2016}
Clack CTM, Alexander A, Choukulkar A, MacDonald AE. Demonstrating the effect of vertical and directional shear for resource mapping of wind power. {\it Wind Energy} 2016\string; 19(9)\string: 1687-1697.
\newblock \href {\doibase https://doi.org/10.1002/we.1944} {doi: https://doi.org/10.1002/we.1944}

\bibitem{Redfern2019}
Redfern S, Olson JB, Lundquist JK, Clack CTM. Incorporation of the Rotor-Equivalent Wind Speed into the Weather Research and Forecasting Model’s Wind Farm Parameterization. {\it Monthly Weather Review} 2019\string; 147(3)\string: 1029 - 1046.
\newblock \href {\doibase https://doi.org/10.1175/MWR-D-18-0194.1} {doi: https://doi.org/10.1175/MWR-D-18-0194.1}

\bibitem{Branlard_Ch7}
Branlard E. Blade Element Theory (BET). In:  Peinke J, Bussel vG. \kern-2pt, eds. {\it Wind Turbine Aerodynamics and Vorticity-Based Methods: Fundamentals and Recent Applications}Cham, Switzerland: Springer International Publishing AG.  2017 (pp. 143--150).

\bibitem{Weick_1930}
Weick FE. The Simply Blade Element Theory. In: New York, New York: McGraw-Hill Book Company.  1930 (pp. 37--50).

\bibitem{Kragh_2014}
Kragh KA, Hansen MH. Load alleviation of wind turbines by yaw misalignment. {\it Wind Energy} 2014\string; 17(7)\string: 971-982.
\newblock \href {\doibase https://doi.org/10.1002/we.1612} {doi: https://doi.org/10.1002/we.1612}

\bibitem{Madsen_2020}
Madsen HA, Larsen TJ, Pirrung GR, Li A, Zahle F. Implementation of the blade element momentum model on a polar grid and its aeroelastic load impact. {\it Wind Energy Science} 2020\string; 5(1)\string: 1--27.
\newblock \href {\doibase 10.5194/wes-5-1-2020} {doi: 10.5194/wes-5-1-2020}

\bibitem{Burton_Ch3}
Burton T, Sharpe D, Jenkins N, Bossanyi E. {\it Aerodynamics of Horizontal-axis Wind Turbines}\string: 41--172; New Jersey, USA: John Wiley \& Sons, Ltd .
\newblock 2001.

\bibitem{Branlard_Ch10}
Branlard E. The Blade Element Momentum (BEM) Method. In:  Peinke J, Bussel vG. \kern-2pt, eds. {\it Wind Turbine Aerodynamics and Vorticity-Based Methods: Fundamentals and Recent Applications}Cham, Switzerland: Springer International Publishing AG.  2017 (pp. 181--214).

\bibitem{Burton_Ed2_Ch3}
Burton T, Jenkins N, Sharpe D, Bossanyi E. {\it Aerodynamics of horizontal axis wind turbines}\string: 39--136; New Jersey, USA: John Wiley \& Sons, Ltd.
\newblock 2~ed. 2011.

\bibitem{Buhl_2005}
Buhl M. A New Empirical Relationship between Thrust Coefficient and Induction Factor for the Turbulent Windmill State. Online;  2005.

\bibitem{akhmatov2007influence}
Akhmatov V. Influence of wind direction on intense power fluctuations in large offshore windfarms in the North Sea. {\it Wind Engineering} 2007\string; 31(1)\string: 59--64.

\bibitem{lindeloew_2009}
Lindel\"ow PJP. Upwind D1. Uncertainties in wind assessment with LIDAR. Online;  2009.

\bibitem{Stull_ch3}
Stull R. {\it Application of the Governing Equations to Turbulent Flow}\string: 75--114; Dordrecht, The Netherlands: Kluwer Academic Publishers .
\newblock 1988.

\bibitem{Wyngaard_ch5}
Wyngaard JC. {\it Turbulent fluxes}\string: 75--88; Cambridge, United Kingdom: Cambridge University Press .
\newblock 2010.

\bibitem{Peinke_Ch11}
Emeis S, T{\"u}rk M. Comparison of Logarithmic Wind Profiles and Power Law Wind Profiles and their Applicability for Offshore Wind Profiles. In:  Peinke J, Schaumann P, Barth S. \kern-2pt, eds. {\it Wind Energy}Proceedings of the Euromech Colloquium. Berlin, Germany: Springer Science+Business Media.  2007 (pp. 91--156).

\bibitem{Aird_2021}
Aird JA, Barthelmie RJ, Shepherd TJ, Pryor SC. WRF-simulated low-level jets over Iowa: characterization and sensitivity studies. {\it Wind Energy Science} 2021\string; 6(4)\string: 1015--1030.
\newblock \href {\doibase 10.5194/wes-6-1015-2021} {doi: 10.5194/wes-6-1015-2021}

\bibitem{Kalverla_2019}
Kalverla PC, Duncan~Jr. JB, Steeneveld GJ, Holtslag AAM. Low-level jets over the North Sea based on ERA5 and observations: together they do better. {\it Wind Energy Science} 2019\string; 4(2)\string: 193--209.
\newblock \href {\doibase 10.5194/wes-4-193-2019} {doi: 10.5194/wes-4-193-2019}

\bibitem{Luiz_2022}
Weide~Luiz E, Fiedler S. Spatiotemporal observations of nocturnal low-level jets and impacts on wind power production. {\it Wind Energy Science} 2022\string; 7(4)\string: 1575--1591.
\newblock \href {\doibase 10.5194/wes-7-1575-2022} {doi: 10.5194/wes-7-1575-2022}

\bibitem{Stull_ch6}
Stull R. {\it Turbulence Closure Techniques}\string: 197--250; Dordrecht, The Netherlands: Kluwer Academic Publishers .
\newblock 1988.

\bibitem{Walter_2007}
Walter KR. {\it Wind power systems in the stable nocturnal boundary layer}. Dissertation. Texas Tech University, Texas, USA;  2007.

\bibitem{Brand_2011}
Brand AJ, Peinke J, Mann J. Turbulence and wind turbines. {\it Journal of Physics: Conference Series} 2011\string; 318(7)\string: 072005.
\newblock \href {\doibase 10.1088/1742-6596/318/7/072005} {doi: 10.1088/1742-6596/318/7/072005}

\end{thebibliography}

\end{document}